\documentclass[11pt]{article}
\PassOptionsToPackage{table}{xcolor}
\usepackage{acl}
\usepackage{times}
\usepackage{amsmath}
\usepackage{amssymb}
\usepackage[T1]{fontenc}
\usepackage[utf8]{inputenc}
\usepackage{microtype}
\usepackage{inconsolata}
\usepackage{graphicx}
\usepackage{booktabs}
\usepackage{makecell}
\usepackage{subcaption}
\usepackage{enumitem}
\usepackage{xcolor}
\usepackage{tcolorbox}\tcbuselibrary{breakable,skins}
\usepackage{pifont}
\usepackage{adjustbox}
\usepackage{tikz}
\usetikzlibrary{arrows.meta,positioning,fit,backgrounds}
\usepackage{algorithm}
\usepackage{algpseudocode}
\usepackage{placeins}
\usepackage{pgfplots}
\pgfplotsset{compat=1.18}

\newcommand{\excise}{\textsc{Excise}}
\newcommand{\xbench}{\textsc{X-Bench}}
\newcommand{\cmark}{\textcolor{green!55!black}{\ding{51}}}
\newcommand{\xmark}{\textcolor{red!70!black}{\ding{55}}}
\newtheorem{proposition}{Proposition}

\definecolor{cOff}{HTML}{7F7F7F}
\definecolor{cDense}{HTML}{0072B2}
\definecolor{cCE}{HTML}{D55E00}
\definecolor{cExcise}{HTML}{009E73}
\newcommand{\oursrow}{\rowcolor{cExcise!8}}

\title{\textsc{EXCISE}: Query-Side Exclusion for Late-Interaction Retrieval}

\author{
  Mohammed Ali \quad Abdelrahman Abdallah \quad Adam Jatowt \\
  University of Innsbruck, Innsbruck, Austria \\
  \texttt{\{mohammed.ali, abdelrahman.abdallah, adam.jatowt\}@uibk.ac.at}
}

\begin{document}
\maketitle

\begin{abstract}
Late-interaction retrievers handle exclusion queries poorly. When a user asks for ``X but not Z'', the additive MaxSim score rewards the tokens that name Z, so documents about Z are promoted rather than suppressed, a failure we call \emph{exclusion inversion}. We show that no method we tested on the frozen token embeddings recovers this constraint; the difficulty lies in identifying the excluded topic, which depends on the query alone. \excise{} therefore confines learning to the query path and leaves the index frozen. Two query-side modules (1.5M parameters in total) detect the excluded topic and re-embed a 100-document shortlist, and a parameter-free rule demotes the candidates that match it. Across six collections and three backbones, \excise{} outperforms each backbone's frozen and fine-tuned baselines on every collection. It raises ExcluIR success@10 from 0.10--0.22 to 0.84--0.85 and Boolean NOT accuracy from 0.25--0.29 to 0.84--0.92. Pooled over 1{,}860 test queries, \excise{} matches the strongest fine-tuned cross-encoder (568M parameters) while preserving the retrieval quality of its frozen index, whereas fine-tuned rerankers degrade ordinary retrieval. To support further work, we release \xbench{}, a tiered benchmark of explicit, cue-free and compound exclusions with no-harm and Boolean controls.\footnote{The code and the benchmark are available at \url{https://github.com/DataScienceUIBK/EXCISE}.}
\end{abstract}

\section{Introduction}
\label{sec:intro}

Retrieval systems are often asked to rule out a topic. A user searching for \emph{electric cars, but not Tesla} should be shown documents about Nissan and not documents about Tesla. Late-interaction retrievers such as ColBERT~\citep{khattab2020colbert} handle the first part of this query correctly and reverse the second: the Tesla page is not merely retrieved but ranked first. Naming the excluded topic is what raises unwanted documents to the top. Across three ColBERT backbones, frozen retrieval solves only $0.10$--$0.22$ of the exclusion queries in ExcluIR~\citep{zhang2024excluir}, and on Reason-ModernColBERT it leaves an excluded document in the top ten for more than 80\% of them. This failure matters most where even a few leaked documents can invalidate the result, as in legal discovery, systematic review and regulatory search.

The scoring rule is the reason. ColBERT keeps one embedding per token and scores a document by summing each query token's strongest match within that document~\citep{khattab2020colbert}. Every token contributes positively, so the tokens naming the excluded topic contribute most to the documents that cover it. We call this \emph{exclusion inversion} and show in \S\ref{sec:nogo} that it follows directly from an identity of MaxSim. As a result, the inversion tends to be strongest on queries that spend the most tokens on the exclusion.

Neither of the two standard remedies is satisfactory. Fine-tuning the encoder teaches exclusion, but the corpus must be re-encoded at every model update, and the adapted encoder usually loses accuracy on queries containing no exclusion. Query-time alternatives avoid re-indexing, yet bring other drawbacks: cross-encoders re-rank every query, whether or not it contains an exclusion, and have no dedicated suppression mechanism, while instruction-following retrievers rely on large instruction-tuned encoders and offer no guarantee that ordinary queries remain unaffected (\S\ref{sec:related}).

To avoid these drawbacks, we first ask which part of the task requires learning. The frozen vectors alone are not enough: a supervised probe can barely separate a wanted document from an unwanted one at any layer. All the nine test-time interventions that we evaluate need the excluded topic as input. Once that topic is supplied, however, a plain demotion rule over the same vectors raises ExcluIR success@10 from $0.135$ to $0.866$. The difficulty therefore lies in identifying the excluded topic from the query alone, not in scoring it, so learning can be confined to the query path.

\excise{} is built on this finding. 
Two lightweight query-side modules, a detector and an exclusion adapter (1.5M parameters against the 149M Reason-ModernColBERT backbone), are trained once offline while the stored index is never modified.
The detector identifies the excluded topic and acts as a switch: when it does not fire, the frozen ranking is returned unchanged. When it fires, the adapter re-embeds the top-100 candidate shortlist and a parameter-free rule lowers each candidate's score by how far its match to the excluded topic stands out from the rest of the shortlist. In the earlier example, documents about Tesla are demoted while those about Nissan rise. Measuring the penalty against the shortlist rather than a fixed threshold also limits the harm of a false firing, in which the detector fires on a query that contains no exclusion. With no real topic to exclude, no candidate stands out from the rest, so the penalty is close to uniform and has little effect on the ranking (\S\ref{sec:results-noharm}).

Across three backbones, \excise{} raises ExcluIR success@10 from $0.10$--$0.22$ to $0.84$--$0.85$, and Boolean NOT accuracy from $0.25$--$0.29$ to $0.84$--$0.92$. Almost every fine-tuned system we compare gains exclusion at a cost to ordinary retrieval, whereas \excise{} preserves the retrieval quality of the frozen index on all three backbones.

Our contributions are:
\begin{enumerate}[nosep,leftmargin=*]
\item \textbf{Exclusion inversion.} We formalise why MaxSim promotes unwanted documents on exclusion queries and show that none of the methods we tested recovers the exclusion constraint from the frozen token vectors; the difficulty lies in identifying the excluded topic from the query (\S\ref{sec:nogo}).
\item \textbf{\excise{}.} We introduce a query-time operator that corrects this failure over a strictly frozen index. It outperforms each backbone's frozen and fine-tuned baselines on every collection. Pooled over all queries at the same input budget, it matches the strongest fine-tuned cross-encoder without degrading ordinary retrieval (\S\ref{sec:method},~\S\ref{sec:results}).
\item \textbf{\xbench{}.} We release a tiered benchmark of explicit, \emph{cue-free} and \emph{compound} exclusions over encyclopedic, financial question-answering, biomedical, sustainability-reporting, financial-regulatory and EU-legal corpora, with no-harm and Boolean controls (\S\ref{sec:bench}).
\end{enumerate}

\section{Related Work}
\label{sec:related}

Neural retrievers handle negation poorly. NevIR~\citep{weller2024nevir} tests retrievers on pairs of documents that differ only by a negation and finds that most perform at chance. ExcluIR~\citep{zhang2024excluir} moves negation from the documents into the query and reports large drops for dense retrievers, consistent with earlier findings that language models mishandle negation at the token level~\citep{kassner2020negation,ettinger2020bert}. A reproduction study~\citep{vandenelsen2025reproducing} shows that only cross-encoders transfer between the two benchmarks, a finding that we reconcile with our results in Appendix~\ref{app:related}. Existing benchmarks usually mark exclusions with a negation cue (e.g., \emph{not} or \emph{without}). QUEST~\citep{malaviya2023quest} tests implicit set operations, including difference, but is limited to entity-seeking queries. \xbench{} adds \emph{cue-free} and \emph{compound} tiers over six domains, together with no-harm and Boolean controls.
 
\looseness=-1 Each existing remedy has a drawback. Lexical filtering~\citep{manning2008ir} relies on a sparse index and an explicit negation cue, so it cannot handle cue-free exclusions. Instruction-following retrievers~\citep{asai2023tart,weller2024promptriever} require large instruction-tuned encoders and offer no guarantee that following an exclusion instruction leaves unrelated queries unaffected, though in our evaluation Promptriever model ~\citep{weller2024promptriever} preserves ordinary retrieval and falls short on exclusion instead. Cross-encoder rerankers~\citep{nogueira2019passage} re-score every query, whether or not it contains an exclusion, and have no dedicated suppression mechanism. Neural-symbolic re-\hspace{0pt}ranking~\citep{xu2025nsir} adds further overhead through a symbolic stage that maps queries and documents into first-order logic. Finally, document-side adaptation, whether a full fine-tune or a LoRA adapter~\citep{hu2022lora}, requires re-encoding the corpus after every model update. \excise{} avoids these drawbacks by confining learning to the query path: it learns to identify the excluded topic instead of matching a cue, re-ranks only when its detector fires, and never re-encodes the corpus. Appendix~\ref{app:related} discusses each family in detail, and \S\ref{sec:results-main} and \S\ref{sec:results-strong} report results for the systems we compare against.


\section{Diagnosing Exclusion Inversion}
\label{sec:nogo}

\paragraph{Task and metric.} Formally, each query $q$ consists of a set of wanted topics $X$ and a set of excluded topics $Z$. The gold document $g$ covers every topic in $X$ and no topic in $Z$. The confusable negative document $n$ covers every topic in $X$ and at least one topic in $Z$, so it resembles the gold document but must not be returned. Let $\mathbf{v}_t$ denote the frozen encoder's $\ell_2$-normalised embedding of token $t$. ColBERT scores a document $d$ by
\begin{equation}\label{eq:maxsim}
S(q,d) = \sum_{t \in q} \max_{u \in d} \, \langle \mathbf{v}_t, \mathbf{v}_u \rangle .
\end{equation}
Our primary metric, \emph{success@10}, enforces the exclusion constraint directly, counting a query as solved only when the gold document appears in the top ten while no excluded document appears there. We report it alongside \emph{hit@10}, which measures whether the gold document reaches the top ten at all, and \emph{leak}, the average fraction of a query's excluded documents left in the top ten.

\paragraph{Exclusion inversion.} The additive form of Eq.~\ref{eq:maxsim} is the source of the difficulty. To see why, we split the query into its wanted part $q_X$ and its excluded part $q_Z$, let $\varepsilon = S(q_X, n) - S(q_X, g)$ be the gap between the negative and the gold documents on the wanted topic, and let $\delta = \bigl(S(q_Z, n) - S(q_Z, g)\bigr) / |q_Z|$ be the negative's mean per-token advantage on the excluded tokens. Because Eq.~\ref{eq:maxsim} is additive over query tokens, the score difference between the two documents follows directly:

\begin{proposition}[Exclusion inversion]\label{prop:mono}
$S(q,n) - S(q,g) = \varepsilon + \vert q_Z \vert \, \delta$.
\end{proposition}

\noindent The right-hand side of Proposition~\ref{prop:mono} has two terms, only one of which stems from the exclusion clause. The first, $\varepsilon$, quantifies the relevance margin on the wanted topic (e.g., \emph{electric cars}): it is positive when the negative document covers the topic better, zero when both are equally relevant, and negative when the gold document is more relevant. The second term captures the contribution of the exclusion tokens. Because the excluded topic appears in the confusable negative, these tokens typically match it more strongly than the gold document by $\delta$ per token on average, giving a total contribution of $\vert q_Z \vert \delta$. Consequently, for fixed $\varepsilon$ and $\delta > 0$, a sufficiently long exclusion clause overrides the gold document's relevance margin, and the document that the clause is meant to demote outscores the gold. The inversion is also common in practice: on the $1{,}860$ test queries, the confusable negative outscores the gold document on 50\% of the longest exclusions and 19\% of the shortest (Appendix~\ref{app:nogo}). This failure is one of suppression rather than retrieval: frozen ColBERT ranks the ExcluIR gold document in the top ten for 94\% of queries, but solves only $0.135$ of them because an excluded document appears there as well.

\paragraph{Limits of the frozen vectors.}
Since the sum in Eq.~\ref{eq:maxsim} causes the failure, a natural question is whether the frozen embeddings themselves could still distinguish the gold document from its confusable negative. Two tests show they cannot. The first trains a supervised probe to classify gold documents against confusable negatives; at four layers spanning the encoder and at the final projection, the probe reaches at most $0.584$ accuracy against a chance level of $0.50$ (Table~\ref{tab:probe}). The second evaluates nine test-time interventions that add no trained parameters to the retrieval path, including subspace orthogonalisation and negative-centroid feedback. All nine share one limitation: each requires the excluded topic as input. The strongest among them additionally requires a sparse index alongside the dense one, yet with a predicted topic it does not outperform the plain demotion rule it replaces (Table~\ref{tab:ninefix}). The underlying reason is identical across all nine. Honouring an exclusion means comparing how strongly a document matches the wanted topic (\emph{electric cars}) against how strongly it matches the excluded one (\emph{Tesla}). MaxSim does not perform this comparison: it sums every query token's best match with the same positive sign, so tokens naming the excluded topic inflate the scores of the documents that cover it.

\paragraph{The missing element is the topic, not the scoring.} Once the gold excluded topic is supplied, a plain MaxSim($Z$) demotion over the same frozen vectors raises ExcluIR success@10 from $0.135$ to $0.866$. Identifying the excluded topic is therefore the main difficulty. With the topic supplied by a detector instead, the same rule reaches $0.818$ (Table~\ref{tab:ninefix}), so the $0.049$ gap represents the remaining cost of identification. Since the topic can be read from the query alone, \S\ref{sec:method} confines the learned components to the query path.

\section{The \excise{} Operator}
\label{sec:method}

\begin{figure*}[t]
\centering
\includegraphics[width=0.9\textwidth]{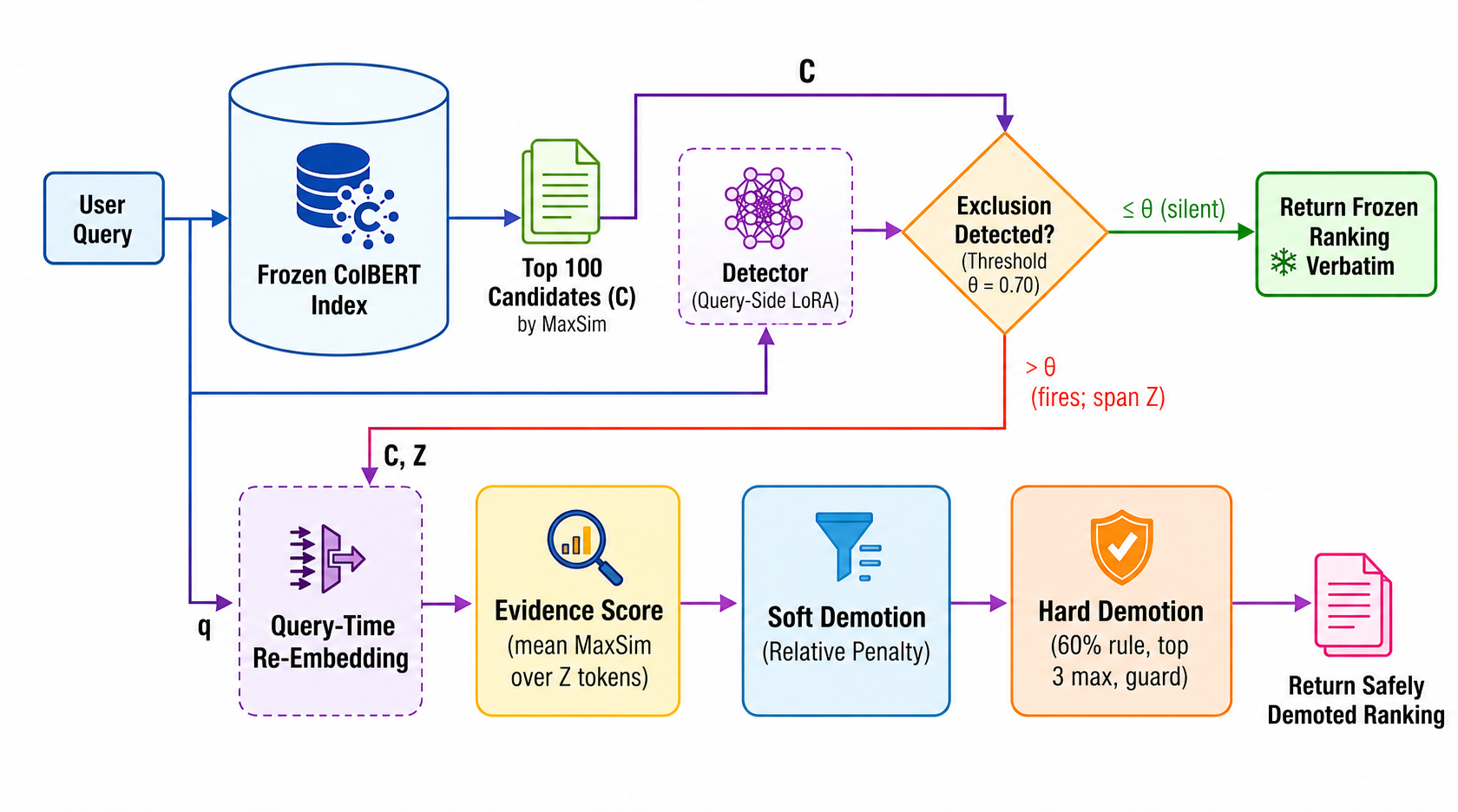}
\caption{\textbf{The \excise{} pipeline at query time.} A third exit, omitted for clarity, returns the re-embedded ranking without demotion when no candidate reaches the evidence floor $\tau$ (\S\ref{sec:method-demotion}).}
\label{fig:arch}
\end{figure*}

\excise{} is a query-side operator that acts on top of a frozen ColBERT index (Figure~\ref{fig:arch}). A query passes through three stages. First, a detector decides whether it contains an exclusion and, if so, marks the excluded topic $Z$. Next, an exclusion adapter re-embeds the query and the top $k = 100$ candidates at query time, without modifying the stored index. Finally, a demotion stage lowers the scores of candidates that match $Z$ and moves the strongest matches to the bottom of the ranking. Only the detector and the exclusion adapter are trained, and together they add 1.5M parameters to the 149M Reason-ModernColBERT backbone. Appendix~\ref{app:example} traces two queries through all three stages end to end.

\subsection{Detection}
\label{sec:method-detector}

The detector is a query-side LoRA module with a token-level span head that tags both the wanted topic $X$ and the excluded topic $Z$ in an IOB scheme, though later stages use only the $Z$ span. It fires when its confidence exceeds $\theta = 0.70$, a single threshold shared across all three backbones (Appendix~\ref{app:detector}). The detector therefore acts as a switch: when it does not fire, \excise{} returns the frozen ranking unchanged, so ordinary queries are affected only by a false firing.

Training uses contrastive minimal pairs that set a conjunctive query (``$A$ and $B$ and $C$'') against an exclusion query, either explicit (``$A$ and $B$ but not $C$'') or cue-free (e.g., ``$A$ and $B$, with $C$ already covered''). The two queries differ only in the exclusion, which forces the detector to learn the exclusion construction rather than to rely on the word \emph{not} (Appendix~\ref{app:detector}).

\subsection{Re-embedding}
\label{sec:method-reembed}

Frozen ColBERT vectors capture what a query is about, not what it rules out. Two documents that differ only in whether they cover the excluded topic also receive similar representations (Table~\ref{tab:probe}). \excise{} re-embeds the query and the shortlist with the exclusion adapter so that such documents can be distinguished more reliably. Because the adapter's training objective includes an exclusion contrast, it learns to separate confusable documents rather than to rank by topic alone.

Re-embedding operates within the shortlist and cannot improve first-stage recall. On \xbench{} this limitation has little effect: the frozen Reason index places the gold document in the top $100$ for at least $98.5\%$ of queries on every collection (Appendix~\ref{app:setup}).

\subsection{Demotion}
\label{sec:method-demotion}

Demotion acts on the re-embedded shortlist. \excise{} first measures how strongly each candidate covers the excluded topic. For every token of the excluded topic, it takes that token's highest similarity to any token of candidate $i$, then averages those values into an evidence score $e_i$. Averaging rather than taking the maximum prevents a single coincidental token match from counting as evidence (Appendix~\ref{app:design}).

A soft penalty then lowers the score of any candidate whose evidence rises above a cutoff $c$:
\begin{equation}\label{eq:demote}
\begin{aligned}
S'(\tilde{q}, \tilde{d}_i) &= S(\tilde{q},\tilde{d}_i) - \lambda M\,\mathrm{ReLU}(e_i - c), \\
c &= \mu(e) + \kappa\,\mathrm{sd}(e),
\end{aligned}
\end{equation}
where tildes mark quantities recomputed by the adapter at query time, $M = \max_j S(\tilde{q},\tilde{d}_j)$ is the largest candidate score, and $\mu(e)$ and $\mathrm{sd}(e)$ are the mean and standard deviation of the evidence scores across the shortlist. The constant $\lambda$ controls the severity of the penalty as a fraction of $M$, while $\kappa$ sets the cutoff: a candidate is penalised only when its evidence exceeds the shortlist mean by more than $\kappa$ standard deviations (Table~\ref{tab:inert}).

The cutoff is relative because, on long queries, most candidates share vocabulary with the excluded topic: evidence scores are then high throughout the shortlist, and a fixed global threshold would penalise almost every candidate, relevant ones included. A cutoff measured against the shortlist's own mean and standard deviation adapts to each query instead. What matters is the relative form of the cutoff and not the width $\kappa$, which can be set anywhere in its grid with no measurable effect on the result (Appendix~\ref{app:sens}).

A relative cutoff alone, however, leaves one case uncovered. Even when no document matches the excluded topic, the candidate with the highest evidence typically exceeds the cutoff and is penalised although no real evidence exists. \excise{} covers this case with an absolute evidence floor $\tau$: when the largest evidence score in the shortlist falls below $\tau$, demotion is disabled and the re-embedded ranking is returned unchanged.

The soft penalty re-ranks the shortlist, but its effect is bounded. A candidate that strongly matches the excluded topic can absorb this penalty and still outrank the gold document if it also matches the wanted topic. The soft penalty alone cannot guarantee that an excluded document leaves the top ten. \excise{} therefore adds a hard demotion step, which moves the candidates that match the excluded topic most strongly to the bottom of the ranking. Several constraints keep this operation safe. First, only candidates whose evidence exceeds the cutoff $c$ and reaches $\phi = 60\%$ of the largest evidence score are eligible. Second, \excise{} demotes at most three of them in decreasing order of evidence, which bounds the harm when the detector misfires. Finally, a guard protects the highest-ranked candidate from hard demotion (Algorithm~\ref{alg:excise}).

The relative cutoff $c$ also keeps the soft penalty safe when the detector misfires.

\begin{proposition}[Flat-penalty invariance]\label{prop:flat}
Let $p_i = \lambda M\,\mathrm{ReLU}(e_i - c)$ be the penalty applied to candidate $i$, let $\Delta$ be the smallest score gap between adjacent candidates in the re-embedded ranking, and let the penalty's spread be $\max_i p_i - \min_i p_i$. If that spread is strictly below $\Delta$, the soft penalty preserves the order of the re-embedded ranking.
\end{proposition}

\noindent When a query carries no real exclusion, the evidence tends to be flat, so the penalty is nearly uniform and its spread is small. By Proposition~\ref{prop:flat}, whenever the spread falls below $\Delta$, the soft penalty leaves the re-embedded ranking unchanged. Harm from the soft penalty therefore depends on its spread, not on the firing itself. \S\ref{sec:results-noharm} measures this harm on the suite where the detector misfires most.

The guarantee covers the soft penalty alone. Hard demotion compares each candidate not only with the cutoff $c$ but also with the largest evidence score in the shortlist. Flat evidence puts every candidate close to that score, so many candidates become eligible for hard demotion (Algorithm~\ref{alg:excise}). The floor $\tau$, the guard and the cap of three bound this case.

\subsection{Cost and Deployment}
\label{sec:method-cost}

The detector runs on every query and adds $12.7$\,ms. When it fires, re-embedding, re-ranking and demotion add a further $15.1$\,ms, so the expected overhead is $12.7 + 15.1\rho$\,ms at firing rate $\rho$, and the worst case at $\rho = 1$ is a total of $1.4\times$ the frozen latency (Appendix~\ref{app:setup}). No generative model is called at query time. A fine-tuned document encoder must re-index the corpus after every model update, whereas \excise{} writes nothing to the index and can be attached or removed without re-indexing.

\section{The \xbench{} Benchmark}
\label{sec:bench}

A method that improves exclusion cannot be judged on exclusion alone: it must also preserve ordinary retrieval and respect Boolean operators. \xbench{} tests all three, providing $1{,}860$ exclusion queries across six collections (Table~\ref{tab:data}). Two collections, ExcluIR~\citep{zhang2024excluir} and FiQA~\citep{maia2018fiqa}, provide in-domain training and test splits with disjoint document pools. The remaining four, TREC-COVID~\citep{voorhees2021treccovid}, ESGenius~\citep{he2025esgenius}, EDGAR-CORPUS~\citep{loukas2021edgar} (EDGAR) and the English subset of MultiEURLEX~\citep{chalkidis2021multieurlex} (EUR-Lex), are held out entirely for zero-shot evaluation. No training query or document appears among the test queries or their gold and negative documents (Appendix~\ref{app:human}). EDGAR, EUR-Lex and ESGenius are indexed at passage level, with medians of $227$, $137$ and $463$ words per unit; the other three collections are indexed as their sources publish them.

\paragraph{Tiers.} \xbench{} introduces three difficulty tiers: the first two appear in short and long variants, and the third is long only. The long variants embed the exclusion mid-sentence. \textbf{T1} uses explicit negation markers such as \emph{not}, \emph{without} or \emph{excluding}. \textbf{T2} contains no negation cue; the exclusion is conveyed through restrictive framing, as in \emph{``With the European heavyweight title already covered, which boxer competed in the light welterweight division at the 2000 Summer Olympics?''}, where \emph{already covered} is the only signal. \textbf{T3} rules out two topics simultaneously, in either a disjunctive form (\emph{neither $A$ nor $B$}) or a mixed form pairing one explicit and one implicit exclusion. A lexical check confirms the separation of T1 from T2: an explicit negation marker appears in all $693$ T1 queries and in none of the $569$ T2 queries (Appendix~\ref{app:human}). Table~\ref{tab:gallery-excl} provides examples.

\paragraph{Construction and quality control.} Queries are generated with GPT-4.1 from mined confusable document pairs, without templates. A query is admitted only if it passes nine gates, grouped into semantic, retrieval and structural checks. The four semantic gates require the excluded topic to be present in the negative document, absent from the gold document, verified by a judge rather than by string matching when the query paraphrases it, and still ruled out by the query read on its own. The two retrieval gates, round-trip and difficulty, require the gold document to be retrievable and the negative to remain competitive with it. The three structural gates reject near-duplicates, require a separate negative for each excluded topic in a compound query, and test for leakage. On an admitted query, a failure therefore largely reflects exclusion rather than retrieval (Appendix~\ref{app:bench}). Round-trip enforces reachability rather than ease: frozen Reason-ModernColBERT passes the gate on every admitted query and still solves only $0.135$ of ExcluIR. As an independent check, a judge from a different model family agrees with $94.4\%$ of admitted labels (Appendix~\ref{app:human}). Human validation is reported in Appendix~\ref{app:humanval}.

\paragraph{Controls and fairness.} Two control suites complement the exclusion tests. The \textbf{no-harm suite}, comprising SciFact~\citep{wadden2020scifact}, NFCorpus~\citep{boteva2016nfcorpus} and ArguAna~\citep{wachsmuth2018arguana} as distributed in BEIR~\citep{thakur2021beir}, measures nDCG@10 on queries containing no exclusions. The \textbf{Boolean suite}, BoolQuestions~\citep{zhang2024boolquestions}, tests conjunction (AND), disjunction (OR) and negation (NOT) separately. Every trainable system, \excise{} and all baselines alike, trains on the same $1{,}900$ retrieval triples for three epochs. Our own adapters use LoRA rank 8, while every external system is trained at the full capacity its architecture allows, and the learning rate of every baseline is selected by the same held-out procedure. \excise{}'s exclusion adapter is additionally trained on $727$ demotion triples from the same training split, which supervise the evidence score of the demotion rule, a component no baseline has; the detector is trained on document-free query labels (Appendix~\ref{app:detector}).

\section{Experimental Setup}
\label{sec:setup}

We evaluate \excise{} on three late-interaction backbones: Reason-ModernColBERT (Reason) and GTE-ModernColBERT-v1 (GTE), both built on ModernBERT~\citep{warner2024modernbert}, and ColBERTv2.0~\citep{santhanam2022colbertv2} (ColBERTv2), built on BERT~\citep{devlin2019bert}. For each backbone, we compare against four baselines: the frozen index, a LoRA fine-tune, a full encoder fine-tune (both re-indexed after adaptation), and a query-only adapter applied to the frozen index. The last baseline isolates learning on the query path from re-indexing the corpus. Beyond these, we include eight external systems (four retrievers and four cross-encoder rerankers), each tested off the shelf and after fine-tuning on the same $1{,}900$ retrieval triples (\S\ref{sec:results-strong}).

Document length is a confound because long documents are split into chunks and scored by their best-matching chunk, and the three backbones differ in how much text they index per document. Reason indexes $10{,}000$ words; GTE and ColBERTv2 index $1{,}000$; external systems, as published, read $512$ tokens. To keep the comparison fair, all methods within a band of Table~\ref{tab:main} use the same document budget, and every external system is re-run at the $10{,}000$-word budget (Table~\ref{tab:strong}). The budgets also make little difference across backbones: only $0.03\%$ of \xbench{} documents exceed $1{,}000$ words, so each backbone indexes almost all documents in full. Chunk size is an encoder property and remains at its native value for each backbone.

All hyperparameters are fixed before any test label is consulted, under identical selection rules for \excise{} and every baseline. Learning rates are chosen by grouped five-fold cross-validation on the training triples. Epochs are fixed at $3$, and our own adapters use LoRA rank $8$, so none is advantaged; the full fine-tunes update every weight and therefore carry more capacity than \excise{}, not less. All operator constants are fixed across all three backbones; their values are given in Table~\ref{tab:inert} and discussed in Appendix~\ref{app:sens}.

We report \textbf{success@10}, \textbf{hit@10} and \textbf{leak} as defined in \S\ref{sec:nogo}. Because success@10 is binary and penalises one leaked document as heavily as nine, we also include xnDCG$_{\beta=1}$, a graded measure that weights a leaked document equally with a missing gold document. On Reason, these two metrics agree in almost all comparisons between \excise{} and a baseline (Appendix~\ref{app:pertier}). Significance is assessed with a paired bootstrap over per-query outcomes ($10{,}000$ resamples, aligned by query ID).

\section{Results}
\label{sec:results}

\begin{table*}[t]\centering
\caption{Exclusion and retrieval results across three ColBERT backbones, each against its own frozen and fine-tuned baselines. The six collection columns
report exclusion success@10; \textbf{No-harm} is mean nDCG@10 on SciFact/NFCorpus/ArguAna and \textbf{NOT} is Boolean-negation pairwise accuracy.
Within each band, \textbf{bold} marks the column maximum and \underline{underline} the strongest baseline; No-harm is read against the frozen
row rather than maximised. Each band uses its own document budget; only $0.03\%$ of documents exceed $1{,}000$ words, so the budgets cannot
separate the bands. $^{\dagger}$Encoder retrained; index must be re-encoded. $^{\ddagger}$Adapter on the \emph{query} alone against the
\emph{frozen} index.}
\label{tab:main}
\resizebox{\textwidth}{!}{%
\begin{tabular}{@{}lccccccc cc@{}}
\toprule
\textbf{System} & \textbf{Re-index?} & \multicolumn{2}{c}{\textbf{In-domain}} & \multicolumn{4}{c}{\textbf{Out-of-domain (zero-shot)}} & \textbf{No-harm} & \textbf{NOT} \\
\cmidrule(lr){3-4}\cmidrule(lr){5-8}\cmidrule(lr){9-9}\cmidrule(lr){10-10}
 & & ExcluIR & FiQA & TREC-C & ESGen & EDGAR & EUR-Lex & (nDCG) & (acc.) \\
\midrule
\rowcolor{gray!12}\multicolumn{10}{c}{\textit{Reason-ModernColBERT, 149M, 10{,}000-word document budget}} \\
Frozen ColBERT & \xmark & 0.135 & 0.183 & 0.187 & 0.170 & 0.303 & 0.281 & 0.5126 & 0.292 \\
LoRA fine-tune$^{\dagger}$ & \cmark & \underline{0.682} & 0.514 & 0.644 & 0.335 & \underline{0.530} & 0.571 & 0.5085 & 0.745 \\
Full fine-tune$^{\dagger}$ & \cmark & 0.581 & \underline{0.543} & 0.618 & 0.354 & 0.501 & 0.475 & 0.4917 & 0.656 \\
Query-only fine-tune$^{\ddagger}$ & \xmark & 0.666 & 0.524 & \underline{0.682} & \underline{0.364} & 0.501 & \underline{0.592} & 0.5046 & 0.741 \\
\oursrow \textbf{\excise{}} (ours) & \xmark & \textbf{0.850} & \textbf{0.755} & \textbf{0.846} & \textbf{0.655} & \textbf{0.677} & \textbf{0.745} & 0.5072 & \textbf{0.891} \\
\midrule
\rowcolor{gray!12}\multicolumn{10}{c}{\textit{GTE-ModernColBERT-v1, 149M, 1{,}000-word document budget}} \\
Frozen ColBERT & \xmark & 0.098 & 0.154 & 0.195 & 0.063 & 0.230 & 0.265 & 0.4911 & 0.292 \\
LoRA fine-tune$^{\dagger}$ & \cmark & 0.673 & 0.505 & \underline{0.709} & 0.296 & \underline{0.513} & \underline{0.612} & 0.5080 & 0.745 \\
Full fine-tune$^{\dagger}$ & \cmark & 0.590 & \underline{0.534} & 0.633 & \underline{0.306} & 0.459 & 0.505 & 0.4684 & 0.692 \\
Query-only fine-tune$^{\ddagger}$ & \xmark & \underline{0.706} & 0.414 & 0.522 & 0.282 & 0.343 & 0.500 & 0.4437 & 0.745 \\
\oursrow \textbf{\excise{}} (ours) & \xmark & \textbf{0.836} & \textbf{0.716} & \textbf{0.808} & \textbf{0.670} & \textbf{0.603} & \textbf{0.699} & 0.4909 & \textbf{0.923} \\
\midrule
\rowcolor{gray!12}\multicolumn{10}{c}{\textit{ColBERTv2.0, 110M, 1{,}000-word document budget}} \\
Frozen ColBERT & \xmark & 0.222 & 0.245 & 0.300 & 0.087 & 0.221 & 0.281 & 0.4663 & 0.251 \\
LoRA fine-tune$^{\dagger}$ & \cmark & \underline{0.646} & \underline{0.505} & \underline{0.644} & 0.248 & \underline{0.479} & \underline{0.597} & 0.4287 & 0.705 \\
Full fine-tune$^{\dagger}$ & \cmark & 0.626 & 0.471 & 0.621 & \underline{0.286} & 0.453 & 0.571 & 0.4075 & 0.664 \\
Query-only fine-tune$^{\ddagger}$ & \xmark & 0.511 & 0.442 & 0.542 & 0.267 & 0.399 & 0.510 & 0.4208 & 0.680 \\
\oursrow \textbf{\excise{}} (ours) & \xmark & \textbf{0.843} & \textbf{0.707} & \textbf{0.735} & \textbf{0.626} & \textbf{0.524} & \textbf{0.617} & 0.4653 & \textbf{0.842} \\
\bottomrule
\end{tabular}}
\end{table*}

\subsection{Exclusion}
\label{sec:results-main}

Table~\ref{tab:main} reports exclusion success@10 for every system on every backbone. Frozen retrieval performs poorly on the task, solving $0.16$--$0.23$ of the pooled queries and leaving the confusable negatives near the top, as Proposition~\ref{prop:mono} predicts. \excise{} is the best system in all eighteen comparisons, each against that backbone's own frozen index and its three fine-tunes, and it never re-encodes the index. Pooled over all six collections, it outperforms the strongest baseline by $0.15$--$0.19$ on each backbone. The smallest per-collection margin, $0.020$, occurs on EUR-Lex with ColBERTv2.

Encoder fine-tuning is less reliable than its in-domain scores suggest. The fine-tuned encoders attain $0.58$--$0.68$ on ExcluIR but fall to $0.25$--$0.35$ on the held-out ESGenius collection. In the pooled comparison, they remain $0.15$--$0.24$ below \excise{} on every backbone, and each requires the corpus to be re-encoded.

\subsection{Ordinary Retrieval and Boolean Control}
\label{sec:results-noharm}

An improvement in exclusion is of limited value if it degrades ordinary retrieval. Table~\ref{tab:noharm} reports both control axes: nDCG@10 on the three no-harm suites, and pairwise accuracy on the Boolean suite. \excise{} matches the frozen index it starts from on every suite, whereas the encoder fine-tunes do not: it loses at most $0.008$ nDCG@10 on any suite, against up to $0.126$ for the fine-tunes on the same suites.

ArguAna is the most demanding of the three, because its long argument-style queries produce the detector's highest false-firing rate even though the suite contains no exclusions: the switch fires on $0.102$ of its queries. One query in ten therefore runs the full operator on a suite with nothing to exclude, and the total cost of doing so is the $0.008$ reported above. This small cost is expected: when no real excluded topic exists, the evidence scores tend to be close to uniform, so the penalty spread is small and the ranking changes little.

On the Boolean suite, the operator acts on negation and leaves the other connectives largely unchanged. Accuracy on NOT rises from $0.25$--$0.29$ to $0.84$--$0.92$, more than tripling on every backbone, while AND and OR fall slightly, by at most $0.048$. Negation is the only connective that the frozen retriever handles poorly and that demotion can address; conjunction and disjunction already score above $0.93$, so there is little room to improve them (Appendix~\ref{app:noharm}).

\subsection{Comparison with External Systems}
\label{sec:results-strong}

\begin{figure}[t]\centering
\begin{tikzpicture}
\begin{axis}[width=\columnwidth,height=5.7cm,
  xlabel={Exclusion success@10},
  ylabel={No-harm nDCG@10},
  xmin=0.16,xmax=0.82,ymin=0.24,ymax=0.60,
  xtick={0.1,0.2,0.3,0.4,0.5,0.6,0.7,0.8},ytick={0.3,0.4,0.5},
  scaled ticks=false,
  xticklabel style={/pgf/number format/fixed,/pgf/number format/precision=1},
  yticklabel style={/pgf/number format/fixed,/pgf/number format/precision=1},
  grid=major,grid style={gray!14,line width=0.3pt},
  axis line style={gray!55},tick style={gray!55},tick align=outside,
  tick label style={font=\footnotesize},
  legend style={font=\scriptsize, at={(0.5,-0.28)}, anchor=north, legend columns=2,
    column sep=10pt, row sep=1pt, draw=none, fill=none},
  legend cell align=left,
  xlabel near ticks,ylabel near ticks,
  every axis x label/.append style={font=\footnotesize},
  every axis y label/.append style={font=\footnotesize},
  clip=false]
\input{latex_tables/tradeoff_points}
\node[anchor=east,font=\scriptsize,cExcise] at (axis cs:0.7731,0.5072) {\textbf{\excise{}}};
\end{axis}
\end{tikzpicture}

\caption{\textbf{Learning exclusion normally costs ordinary retrieval.} Pooled exclusion success@10 over all $1{,}860$ queries against mean no-harm nDCG@10 on SciFact/NFCorpus/ArguAna, every system at the same $10{,}000$-word budget. \S\ref{sec:results-strong} discusses the result.}
\label{fig:tradeoff}
\end{figure}

Table~\ref{tab:strong} compares \excise{} on Reason with eight external systems at the same $10{,}000$-word budget: three dense retrievers, an instruction-following retriever and four cross-encoder rerankers, each evaluated off the shelf and after fine-tuning on our triples.

Off the shelf, none of them solves exclusion. Models from 22M to 1.5B parameters solve at most one ExcluIR query in five, and the largest solves fewer than the smallest, indicating that the failure is structural rather than a matter of capacity. The instruction-following retriever performs better, but requires a 7B model and a separate index.

Fine-tuning improves exclusion for every system, but usually at the cost of ordinary retrieval (Figure~\ref{fig:tradeoff}). Two systems avoid that cost. Promptriever, the 7B instruction-following retriever, is one, gaining $0.033$ nDCG@10, but it trails \excise{} on exclusion across all six collections, by $0.136$ pooled. \excise{} is the other, and the only one that preserves ordinary retrieval while matching the strongest reranker on exclusion.

The input budget also affects the comparison. As published, these systems read only the first $512$ tokens, while at the $10{,}000$-word budget each document is divided into $350$-word passages and scored by the best of them. On ESGenius, whose units have a median length of $463$ words, the best-scoring passage of an excluded document can then be one in which the excluded topic does not appear: the strongest cross-encoder's leak rises by $0.129$ and its lead over \excise{} narrows from $0.146$ to $0.015$. On the other five collections, its success@10 changes by at most $0.010$ (Table~\ref{tab:windowing}; Appendix~\ref{app:strong} isolates the mechanism), and \excise{} indexes the whole document under both conditions.

Pooled over the full query set at this budget, \excise{} matches the strongest of the fine-tuned cross-encoders and leads the other three by $0.079$--$0.198$. Most collections hold too few queries to test on their own, so significance is assessed on the pool. Against the 568M reranker, the difference is $-0.012$, with a $95\%$ confidence interval of $[-0.034, +0.010]$ and $p = 0.277$, so the test does not separate them. They are separated only on ExcluIR, where the reranker is ahead by $0.034$ ($p = 0.024$). The reranker reaches parity on the pool with 568M fine-tuned parameters against \excise{}'s 1.5M, and with about fifteen times the loss in ordinary retrieval, $0.080$ nDCG@10 against $0.005$.

\subsection{Component Analysis}
\label{sec:results-ablation}

Table~\ref{tab:ablation} reports the contribution of each component of \excise{} on Reason. Frozen ColBERT solves $0.135$ of ExcluIR. Applying \excise{}'s soft penalty to the frozen vectors, without re-embedding, raises this to $0.812$, and the remaining stages add $0.038$, bringing the full operator to $0.850$. That a single rule accounts for most of the gain is what \S\ref{sec:nogo} predicts: once the excluded topic is supplied, scoring it is straightforward.

A leave-one-out analysis confirms the role of each component. Dropping the soft penalty while keeping hard demotion lowers success@10 on five of the six collections, by up to $0.066$. Removing the switch forces re-embedding on every query, so the overhead is paid on all traffic rather than on a fraction of it. Finally, removing demotion altogether lowers success@10 by up to $0.320$, more on every collection than removing re-embedding, confirming that demotion carries the exclusion signal (\S\ref{sec:nogo}).

The remaining errors are structural rather than accidental. Gold documents frequently overlap with excluded topics in vocabulary, especially on the longer collections. Because of this overlap, it matters how the token similarities are combined: replacing the average with the maximum lowers success@10 on all six collections, by $0.175$ on ESGenius and $0.087$ on EUR-Lex (Table~\ref{tab:design}). Averaging is what prevents a coincidental match from demoting a valid answer.

\FloatBarrier


\section{Conclusion}
Late-interaction retrievers fail on exclusion queries for a structural reason: additive scoring rewards the tokens that name the excluded topic, so the exclusion itself promotes the documents it is meant to suppress. What resists a rule-based remedy is identifying the excluded topic, which depends on the query alone. \excise{} therefore confines learning to two small query-side modules and leaves the index frozen. Across three backbones, ExcluIR success@10 rises from $0.10$--$0.22$ to $0.84$--$0.85$ and Boolean NOT accuracy from $0.25$--$0.29$ to $0.84$--$0.92$. Among all systems we evaluate, \excise{} on Reason-ModernColBERT is the only one that matches the strongest fine-tuned reranker on exclusion while preserving ordinary retrieval and leaving the index unchanged. We release \xbench{}, a benchmark of explicit, cue-free and compound exclusions, to support further work on this problem.

\section*{Limitations}

Three limitations bound our results. First, \excise{} cannot recover a gold document that its frozen first stage never retrieves, so its ceiling is the shortlist's recall; on \xbench{} that ceiling is at least $0.985$ on every collection, but a corpus with a weaker first stage would bound it lower. Second, exclusion inversion is derived from an identity of MaxSim, so both the diagnosis and the repair are stated for late-interaction scoring and do not transfer unchanged to single-vector or generative scorers, the latter of which we do not compare against. Third, \xbench{} is model-generated and English only; the gates are strict and independent agreement is high, but generation artefacts and cross-lingual transfer remain untested.

\section*{Acknowledgments}

The authors would like to acknowledge the financial support provided by the Austrian Research Agency (FFG) for the project ``AI Enabled Sustainability Jurisdiction Demonstrator'' (project No. 915229). The computational results presented in this work have been achieved using the MUSICA cluster, part of the Austrian Scientific Computing (ASC) infrastructure.

\bibliography{references}

\clearpage
\appendix

\noindent\textbf{How this appendix is organised.} Each part expands one section of the paper.

\begin{itemize}[nosep,leftmargin=2.6em,itemsep=1pt]
\item[\textbf{\ref{app:partA}}] (\S\ref{sec:nogo}) how the inversion grows with exclusion length, the linear probe, and the
nine test-time interventions under oracle and detector $Z$.
\item[\textbf{\ref{app:partB}}] (\S\ref{sec:method}) the operator in full, detector training and calibration, how every
constant was fixed, the design alternatives we tested, and two worked examples.
\item[\textbf{\ref{app:partC}}] (\S\ref{sec:bench}) benchmark construction with the prompts, the nine admission
gates, three validity measurements, the human validation, and a query gallery.
\item[\textbf{\ref{app:partD}}] (\S\ref{sec:setup}) configuration and query-time cost, and the external baselines at both
input budgets.
\item[\textbf{\ref{app:partE}}] (\S\ref{sec:results}) the control suites in full, the component ablation, and
per-tier results.
\item[\textbf{\ref{app:partF}}] (\S\ref{sec:related}) every related family in full, with the costs of each.
\end{itemize}

\section{Evidence Behind \S\ref{sec:nogo}}
\label{app:partA}

\subsection{How the Inversion Grows}
\label{app:nogo}

\noindent\textbf{Proof of Proposition~\ref{prop:mono}.} Eq.~\ref{eq:maxsim} sums over query tokens, and $q_X$ and $q_Z$ partition them, so
$S(q,d) = S(q_X,d) + S(q_Z,d)$ for every document $d$. The difference between the two documents' scores is
therefore $\bigl(S(q_X,n) - S(q_X,g)\bigr) + \bigl(S(q_Z,n) - S(q_Z,g)\bigr)$, which is
$\varepsilon + \vert q_Z \vert \delta$ by the definitions of $\varepsilon$ and $\delta$.

Proposition~\ref{prop:mono} is an algebraic identity: it states what the inversion must do but not how often it
occurs or how large it becomes. This section measures both on all $1{,}860$ test queries with the frozen Reason
backbone.

The inversion occurs on about a third of the queries. The gap between the two documents is $S(q,n) - S(q,g)$,
the amount by which the confusable negative outscores the gold document. It is positive on 32.5\% of the queries, so the
document that the user asked to exclude outranks the wanted one before any operator is applied. This figure
describes admitted pairs rather than document pairs in general, because the negatives were mined to be
confusable (Appendix~\ref{app:bench}).

The gap also grows with the length of the exclusion. Proposition~\ref{prop:mono} predicts that it increases with
the number of query tokens spent on the exclusion. Grouping the queries by that number gives the following mean
gap and inversion rate:

\begin{center}\small
\begin{tabular}{@{}lcccc@{}}
\toprule
excluded tokens & 1--3 & 4--6 & 7--10 & $\ge 11$ \\
\midrule
mean gap & $-2.065$ & $-1.608$ & $-1.110$ & $-0.217$ \\
inversion rate & 19\% & 29\% & 37\% & 50\% \\
\bottomrule
\end{tabular}
\end{center}

\noindent Both the mean gap and the inversion rate increase monotonically across the four groups, as
Proposition~\ref{prop:mono} predicts. Two caveats apply. The token count is a noisy proxy, because cue-free and
compound queries paraphrase the excluded topic instead of naming it directly, and four groups establish a
direction rather than a rate.

\subsection{The Linear Probe}
\label{app:probe}

Proposition~\ref{prop:mono} and the measurements above concern the score. A score may, however, discard
information that the representation still contains, so we next ask whether the frozen vectors separate the two
documents even where the score does not. We call any rule that attempts to recover the constraint from those
vectors a \emph{readout}; the simplest readout is a supervised probe.

We train a logistic regression to distinguish gold documents from confusable negatives, using the frozen token
vectors at four layers spanning the encoder and at the final projection. The probe separates them only
marginally: its accuracy peaks at $0.584$ against a chance level of $0.50$, far below the separation that a
ranking decision would require.

\begin{table}[ht]\centering\small

\caption{A supervised probe cannot usefully separate gold documents from confusable negatives at any of the four layers or at the final projection (chance $= 0.50$; final $=$ the 128-dimensional projection). \textbf{Scope:} the probe is fit on gold/confusable-negative document pairs mined from ExcluIR with document-level splits, and it characterises the \emph{frozen representation} of Reason-ModernColBERT. It is a property of that representation rather than a measurement on any evaluation set.}
\label{tab:probe}
\begin{tabular}{@{}lccccc@{}}
\toprule
Layer & L5 & L11 & L16 & L22 & final \\
\midrule
Probe acc. & 0.551 & 0.576 & 0.584 & 0.581 & 0.539 \\
\bottomrule
\end{tabular}
\end{table}

\subsection{Nine Test-Time Interventions}
\label{app:ninefix}

The supervised probe fails, but a hand-designed readout might succeed, so we evaluate nine. Each replaces the
plain MaxSim($Z$) demotion and is compared with it on ExcluIR and FiQA.

The reference, which we call the \emph{plain} rule, subtracts a candidate's MaxSim($Z$) from its base score
after normalising both over the shortlist, with no cutoff. We use the simplest rule that takes the excluded
topic as input, because the question is whether any readout can substitute for knowing that topic. \excise{} scores above every other row in all four columns. The difference between its two entries on ExcluIR, $0.8899$ against $0.8502$,
gives the cost of identifying the excluded topic instead of receiving it: $0.040$, against $0.049$ for the
reference rule (\S\ref{sec:nogo}).

\begin{table}[t]\centering
\caption{\textbf{Nine test-time interventions, each \emph{replacing} the plain MaxSim($Z$) demotion} and compared with it on the $554$ ExcluIR and $208$ FiQA test queries. In the \textbf{oracle} columns, every row, including \excise{}, receives the excluded topic recorded with the query. In the \textbf{detector} columns, every row receives the topic predicted by the detector; only this setting is deployable. None of the interventions adds a trained parameter to the retrieval path; the learned readout gate trains only a readout of the frozen vectors. \textbf{Bold} marks the best of the nine in each column and, separately, the \excise{} row. A dash marks a value that was not measured.}
\label{tab:ninefix}
\resizebox{\columnwidth}{!}{%
\begin{tabular}{@{}lcccc@{}}
\toprule
 & \multicolumn{2}{c}{\textbf{oracle} $\boldsymbol{Z}$} & \multicolumn{2}{c}{\textbf{detector} $\boldsymbol{Z}$} \\
\cmidrule(lr){2-3}\cmidrule(lr){4-5}
\textbf{Intervention} & ExcluIR & FiQA & ExcluIR & FiQA \\
\midrule
\emph{MaxSim($Z$) demotion, reference} & \emph{0.8664} & \emph{0.7452} & \emph{0.8177} & \emph{0.7067} \\
\midrule
Sparse BM25-$Z$ demotion & \textbf{0.8520} & 0.7308 & \textbf{0.8159} & 0.6779 \\
Positional--lexical (frequency) & 0.8375 & 0.6731 & 0.8051 & 0.6250 \\
Positional--lexical (first position) & 0.8213 & 0.7019 & 0.7978 & 0.6538 \\
Contrastive late-interaction & 0.8069 & \textbf{0.7692} & 0.6715 & \textbf{0.6875} \\
Negative-centroid feedback & 0.7545 & 0.6154 & 0.7256 & 0.5913 \\
Subspace orthogonalisation & 0.6570 & 0.5721 & 0.6661 & 0.5769 \\
Token dropping alone & 0.5722 & 0.4712 & 0.5632 & 0.5000 \\
Learned readout gate & 0.3069 & 0.2981 & 0.3069 & 0.2933 \\
$Z$-conditional re-embedding & 0.1895 & 0.2740 & 0.1895 & 0.2644 \\
\midrule
\emph{frozen retrieval} & \emph{0.1354} & \emph{0.1827} & -- & -- \\
\oursrow \textbf{\excise{}} (ours) & \textbf{0.8899} & \textbf{0.7981} & \textbf{0.8502} & \textbf{0.7548} \\
\bottomrule
\end{tabular}}
\end{table}

Three points limit how the table should be interpreted. First, each intervention changes a single aspect of
the plain rule, whereas \excise{} is the complete operator, combining re-embedding, the relative cutoff and hard demotion. Second, the oracle-$Z$ columns supply a topic that a deployed system would have to identify, so these
results are diagnostic rather than achievable. Third, the study covers only ExcluIR and FiQA, the two
in-domain collections on which the confusable pairs were mined; the four held-out collections are not
included. The nine interventions fall into three families according to what they act on.

\paragraph{Lexical rules.} The first family operates on words. \textbf{Sparse BM25-$Z$ demotion} scores the
excluded topic against the text of the candidate with BM25 over a sparse index and subtracts that score.
\textbf{Positional--lexical} weights each excluded term by its inverse document frequency and has two forms: one scores how early the document first names the topic, and the other how often it names it. \textbf{Token dropping} is the simplest variant: it deletes the excluded tokens from the
query before scoring.

\paragraph{Geometric rules.} The second family operates on vectors. \textbf{Contrastive late-interaction}
scores a candidate on the wanted part of the query and subtracts its match to the excluded topic, each as a
length-normalised MaxSim. \textbf{Subspace orthogonalisation} removes the principal direction of the excluded
topic from the query and document token vectors and re-scores. \textbf{Negative-centroid feedback} follows
Rocchio~\citep{rocchio1971relevance}: it averages the token vectors of the three shortlisted candidates closest
to the excluded topic into a negative centroid, subtracts it from the query tokens, and re-scores. The
\textbf{learned readout gate} is a logistic probe on the pooled frozen document vectors, trained to separate
gold from negative documents. It is the only intervention that is trained, and what it trains is a readout of
the frozen vectors, not a part of the retrieval path.

\paragraph{An encoding-time rule.} The last intervention changes how the document is encoded.
\textbf{$Z$-conditional re-embedding} re-encodes a flagged candidate as the excluded topic followed by the
document, so that the document tokens attend to that topic during encoding. Apart from the readout gate, none
of these rules learns a weight. In this sense the interventions add no trained parameter to the retrieval path.

\paragraph{With an oracle topic.} Given the gold excluded topic, only one intervention exceeds the reference,
and only on one collection: contrastive late-interaction reaches $0.7692$ on FiQA against $0.7452$ for the
reference. The oracle setting is favourable by construction: the gates require the excluded topic to appear in
the negative and not in the gold, so a rule that receives that topic partly recovers the construction label.

\paragraph{With a detector's topic.} A deployable system has only the topic predicted by a detector. With that
topic, the strongest of the nine on ExcluIR, sparse BM25-$Z$ demotion, falls by $0.036$ to $0.8159$. The
decrease is small because the span head predicts $Z$ accurately, so a simple lexical rule retains most of the
benefit; it still does not outperform the reference on either collection. It also requires a sparse index
maintained alongside the dense one, a cost that success@10 does not show. The comparison therefore depends on
identifying the excluded topic rather than on scoring it.

\paragraph{Why the last row fails.} $Z$-conditional re-embedding prepends the excluded topic to each candidate
before encoding. Because the encoder is bidirectional, the prepended topic affects every token of the document,
including those of the gold. Prepending raises the gold's match to the excluded topic by $0.142$, but the
negative's by only $0.045$. The gap between the two therefore falls from $0.145$ to $0.048$, and the penalty
that the rule assigns is larger for the negative than for the gold on only $44\%$ of queries.

\section{Method Details}
\label{app:partB}

\subsection{The Operator in Full}
\label{app:algo}

\begin{algorithm}[h]
\small
\caption{\excise{} at query time (index frozen)}
\label{alg:excise}
\begin{algorithmic}[1]
\State $C \gets \textsc{FrozenTopK}(q)$ \Comment{first stage, index untouched}
\State $Z \gets \textsc{Detector}(q)$ \Comment{fires iff score $> \theta$}
\If{$Z = \varnothing$} \Comment{switch silent}
  \State \Return $C$ \Comment{frozen order, verbatim}
\EndIf
\State re-embed $q$, $Z$ and each $d_i \in C$ with the adapter
\State $\tilde{S}(d_i) \gets S(\tilde{q}, \tilde{d_i})$ \Comment{re-rank $C$}
\State $e_i \gets$ mean over $t \in Z$ of $\max_{u \in \tilde{d_i}}\langle \tilde{\mathbf{v}}_t, \tilde{\mathbf{v}}_u\rangle$
\If{$\max_i e_i < \tau$} \Comment{no real evidence anywhere}
  \State \Return $C$ sorted by $\tilde{S}$ \Comment{re-ranked, undemoted}
\EndIf
\State $c \gets \mu(e) + \kappa\,\mathrm{sd}(e)$ \textbf{if} $|C| \ge 4$ \textbf{else} the absolute rule
\State $S'(d_i) \gets \tilde{S}(d_i) - \lambda M\cdot\mathrm{ReLU}(e_i - c)$ \Comment{Eq.~\ref{eq:demote}}
\If{$|C| \ge 4$}
  \State $D \gets \{\,d_i : e_i \ge \phi\cdot\max_j e_j \text{ and } e_i > c\,\}$, keep the $3$ largest
  \State $D \gets D \setminus \{d^\ast\}$, where $d^\ast = \arg\max_i \tilde{S}(d_i)$
\Else
  \State $D \gets \varnothing$ \Comment{the soft penalty at line 13 suffices}
\EndIf
\State $S'(d) \gets -\infty$ \textbf{for} $d \in D$ \Comment{hard demotion}
\State \Return $C$ sorted by $S'$
\end{algorithmic}
\end{algorithm}

\noindent Two lines of the algorithm require explanation. Line 12 falls back to what the algorithm calls
\emph{the absolute rule}, which measures the penalty against a fixed threshold instead of the spread of the
shortlist. The cutoff is then $c = 0$ instead of $\mu(e) + \kappa\,\mathrm{sd}(e)$, each candidate is penalised
in proportion to its own evidence, and the floor $\tau$ is the only gate. The operator uses this rule only when
the shortlist holds fewer than four candidates (Appendix~\ref{app:sens}). All Boolean-suite queries fall into
this case, because their candidate sets hold two or three documents.

\noindent Line 16 is the guard from \S\ref{sec:method-demotion}. It removes from the set $D$ the candidate that is ranked first after re-embedding, so hard demotion cannot move this candidate to the bottom of the ranking. The protected candidate can be the confusable negative itself. In that case the soft penalty of line 13 still applies, so the negative is penalised even though it is not moved to the bottom.

\noindent\textbf{Proof of Proposition~\ref{prop:flat}.} Let $i$ and $j$ be adjacent in the re-embedded ranking,
so that $\tilde{S}(i) - \tilde{S}(j) \ge \Delta$. After the penalty their difference is
$\bigl(\tilde{S}(i) - \tilde{S}(j)\bigr) + (p_j - p_i)$, which is at least
$\Delta - \bigl(\max_k p_k - \min_k p_k\bigr)$ and therefore positive whenever the spread falls below $\Delta$.
Every adjacent pair keeps its order, so the ranking is preserved.

\subsection{Detector Training and Calibration}
\label{app:detector}

The detector (line 2 of Algorithm~\ref{alg:excise}) is a query-side LoRA adapter (rank 8, last six layers) with a
linear token head over the per-token representations of the query. The head tags each query token in an IOB
scheme over the wanted topic $X$ and the excluded topic $Z$. The \emph{span score} of the detector is the
maximum excluded-topic probability over the content tokens of the query, and the operator fires when this
score exceeds $\theta$.

The training data are document-free classification labels drawn from the training splits of ExcluIR and FiQA and from Quora, SciDocs~\citep{cohan2020specter}, FEVER~\citep{thorne2018fever}, NQ~\citep{kwiatkowski2019natural} and DBpedia~\citep{hasibi2017dbpedia} queries, as distributed in BEIR~\citep{thakur2021beir}; none of these queries appears in any test or control set. No ArguAna query is used for training.

\paragraph{Contrastive pairs.} A detector can reach high accuracy through shortcuts, and the training set is
constructed to remove them. Its main device is the \emph{contrastive minimal pair}: for every topic frame, an
exclusion query is paired with a conjunction over the same topics, ``$A$ and $B$ and $C$''. The exclusion
query is either explicit, ``$A$ and $B$ but not $C$'', or cue-free, ``$A$ and $B$, with $C$ already covered''.
The two queries share their topics and differ only in whether an exclusion is expressed, so neither the topic
nor the word \emph{not} is informative about the label. The positive spans are the content tokens of the
excluded topic, excluding any token that also belongs to the wanted topic, which further discourages the head
from firing on the wanted topic.

\paragraph{Length balancing.} Minimal pairs control for topic and cue but not for length. In natural sources,
exclusion queries are longer than ordinary ones, so length can serve as a proxy for the label. A detector that
has learned to fire on long queries misfires on every long ordinary query, and firing on a query without an
exclusion is the source of the operator's potential harm. We therefore balance the training set over the full
$2\times2$ grid of query length and fire/no-fire label.

Table~\ref{tab:detector} shows the effect of the balancing. All rows are measured in the same setting, so they
should be compared with one another rather than with the deployed results reported elsewhere. With full
balancing, the detector is equally sensitive at both lengths, with a fire recall of $0.993$ on long queries and
$0.987$ on short ones. Adding only long \emph{no-fire} examples introduces the opposite shortcut: recall is
$0.922$ on long queries against $0.960$ on short ones, and in-domain exclusion on FiQA falls from $0.755$ to
$0.654$. Without either addition the original shortcut remains, and no threshold compensates for it. At
$\theta = 0.70$ this detector fires on $0.996$ of exclusion queries and on $0.240$ of ordinary ones; the only
threshold within the false-firing budget fires on no query, so the operator never runs and exclusion equals the
$0.135$ of frozen retrieval on ExcluIR.

\begin{table}[ht]\centering\small

\caption{\textbf{Effect of balancing the detector's training data over query length}, on Reason-ModernColBERT. Each row is evaluated at its own threshold $\theta$. All three thresholds meet the $1\%$ false-firing budget on SciFact and NFCorpus queries, and the three ArguAna scores differ by at most $0.008$. The last row is the deployed detector.}
\label{tab:detector}
\resizebox{\columnwidth}{!}{%
\begin{tabular}{@{}lccccccc@{}}
\toprule
& & \multicolumn{2}{c}{\textbf{fire recall}} & \textbf{false firing} & \textbf{ArguAna} & \multicolumn{2}{c}{\textbf{succ@10}} \\
\cmidrule(lr){3-4}\cmidrule(lr){7-8}
\textbf{Detector training data} & $\theta$ & long & short & $\downarrow$ & nDCG & ExcluIR & FiQA \\
\midrule
base: firing rows only, $6{,}511$ & 0.98 & 0.000 & 0.000 & 0.000 & 0.405 & 0.135 & 0.183 \\
$+$ long no-fire, $10{,}200$ & 0.94 & 0.922 & 0.960 & 0.005 & 0.405 & 0.850 & 0.654 \\
$+$ long fire \emph{and} no-fire, $12{,}871$ & 0.70 & 0.993 & 0.987 & 0.009 & 0.397 & 0.850 & 0.755 \\
\bottomrule
\end{tabular}}
\end{table}

\paragraph{The firing threshold.} The remaining calibration decision is the value of $\theta$. A threshold is
admissible if its false-firing rate on the ordinary SciFact and NFCorpus queries stays within the budget. This check is based on the query text alone and uses no relevance labels.
Among admissible thresholds we apply the noise test of Appendix~\ref{app:sens}, with fire recall measured on
training-split exclusion queries only. Youden's $J$ (sensitivity plus specificity minus one) selects a
different threshold on each backbone: $0.85$, $0.80$ and $0.10$. We keep $\theta = 0.70$ for all three and
test this choice end to end. Raising the threshold on Reason to $0.85$ changes ESGenius from $0.655$ to
$0.631$ and EUR-Lex from $0.745$ to $0.760$, in opposite directions, and a paired bootstrap finds neither
difference significant ($p = 0.117$ and $p = 0.095$). The threshold $\theta = 0.70$ meets the false-firing budget on all three backbones, so every backbone is evaluated with the same threshold.

\subsection{How the Operator Constants Were Set}
\label{app:sens}

The operator has six constants: the firing threshold $\theta$, the evidence floor $\tau$, the cutoff width $\kappa$, the penalty scale $\lambda$, the hard-demotion fraction $\phi$ and the hard-demotion cap. Each is swept on the
training split only ($1{,}100$ queries: $700$ from ExcluIR and $400$ from FiQA, with no overlap with any test
set), and each is subjected to the same test before adoption.

\paragraph{The noise test.} A grid search always returns a best value for a constant, but that choice is informative only when the objective varies across the grid by more than its sampling error. When the overall spread falls below this error, the ranking of the tested values may not hold on another sample. We call such a constant \emph{inert} and therefore retain the default fixed before the search (Table~\ref{tab:inert}).

None of the six constants has a measurable effect within its grid: five are inert, and the evidence floor is never triggered on the training split. The exclusion results therefore do not depend on the particular values
chosen: the operator reaches $0.850$ on ExcluIR with every constant at its pre-specified default, and varying
any single constant across its admissible grid changes the result by less than the sampling error of that
grid. The operator is sensitive to the \emph{form} of its rules rather than to the values inside them;
Appendix~\ref{app:design} examines the form by replacing the rules themselves.

\begin{table}[ht]\centering
\caption{\textbf{The six operator constants and the verdict of the noise test.} Each constant is swept on the training split alone and adopted only if the objective's spread across its whole grid exceeds its own sampling error: \emph{active} if it does, \emph{inert} if it does not, and \emph{not triggered} if the constant never takes effect on that split. Values are the deployed settings. Appendix~\ref{app:sens} gives the protocol and discusses each row.}
\label{tab:inert}
\small
\resizebox{\columnwidth}{!}{%
\begin{tabular}{@{}lllp{4.2cm}@{}}
\toprule
\textbf{Constant} & \textbf{Value} & \textbf{Verdict} & \textbf{Evidence} \\
\midrule
firing threshold $\theta$ & $0.70$ & inert & Youden's $J$ at $0.70$ is within $0.007$ of its maximum on all three backbones; $0.70$ meets the false-firing budget on each \\
evidence floor $\tau$ & $0.35$ & not triggered & every training query is a genuine exclusion with evidence above the floor; only false firings trigger it, and these occur in the no-harm suites, whose results are not used for tuning \\
penalty scale $\lambda$ & $0.97$ & inert & spread of $0.001$, $0.008$ and $0.004$ on the three backbones across $0.60$/$0.97$/$1.30$, inside sampling error \\
hard-demotion fraction $\phi$ & $0.6$ & inert & $0.80$/$0.85$/$0.90$ within $0.006$ of the deployed $0.6$ on every test collection, in a later check (Table~\ref{tab:design}) \\
hard-demotion cap & $3$ & inert & capping at one moves no test collection by more than $0.010$, in the same check (Table~\ref{tab:design}) \\
cutoff width $\kappa$ & $0.5$ & inert & spread $0.010$ on FiQA-train and $0.009$ on ExcluIR-train, both inside their own binomial noise (Table~\ref{tab:sigma}) \\
\bottomrule
\end{tabular}}
\end{table}

\paragraph{The cutoff width $\kappa$.} A pooled sweep can hide an effect that is present in only one training
source. We therefore sweep $\kappa$ on each source separately, on the training split only
(Table~\ref{tab:sigma}). It is inert on both sources and on the pool: the spread is $0.0085$ on ExcluIR-train
against a noise floor of $0.0286$, $0.0100$ on FiQA-train against $0.0382$, and $0.0017$ pooled against
$0.0229$.

\begin{table}[t]\centering
\caption{\textbf{Cutoff width $\kappa$ swept on each training source separately.} Train-split success@10 at each value, with the spread against that source's own binomial noise floor. No source's spread exceeds its floor, so $\kappa$ is inert on both and on the pool. The sweep runs at $0.25$, $0.5$ and $1.0$; the deployed value is the middle of the three.}
\label{tab:sigma}
\small
\resizebox{\columnwidth}{!}{%
\begin{tabular}{@{}lccc@{}}
\toprule
$\boldsymbol{\kappa}$ & \textbf{ExcluIR-train} & \textbf{FiQA-train} & \textbf{pooled} \\
 & \textit{n}=700 & \textit{n}=400 & \textit{n}=1{,}100 \\
\midrule
0.25 & 0.8229 & 0.8250 & 0.8237 \\
\textbf{0.5} \textit{(deployed)} & 0.8257 & 0.8225 & 0.8245 \\
1.0 & 0.8314 & 0.8150 & 0.8254 \\
\midrule
spread & 0.0085 & 0.0100 & 0.0017 \\
2\,$\times$\,binomial SE & 0.0286 & 0.0382 & 0.0229 \\
verdict & inert & inert & inert \\
\bottomrule
\end{tabular}}
\end{table}

The deployed value, $\kappa = 0.5$, is the middle setting on every source and lies $0.0009$ below the best
pooled value. The two sources also rank the settings in opposite orders, which is consistent with a constant
that has no effect.

\paragraph{The evidence floor $\tau$.} The floor disables the penalty when no candidate carries substantial
evidence, which is characteristic of a false firing. It cannot be calibrated on the training split, because
every query there is a genuine exclusion whose evidence lies above the floor, so the result does not change
for settings from $0.25$ to $0.45$. The floor is triggered only by false firings, which occur in the no-harm suites; their results are reported but never used for tuning. We therefore retain $\tau = 0.35$
on the basis of its function rather than a sweep.

\paragraph{The penalty scale $\lambda$.} A stronger penalty trades relevance for exclusion, so $\lambda$ sets
an operating point rather than having an optimum. We keep $\lambda = 0.97$, the default fixed before the sweep.

\paragraph{Small shortlists.} The relative cutoff estimates a mean and a standard deviation from the shortlist,
which requires a minimum number of candidates. With two candidates the sample deviation equals
$|a-b|/\sqrt{2}$, a function of the single score gap, so the relative cutoff reduces to a rescaled absolute rule.
In that case $\kappa$ would also act as a switch, suppressing demotion entirely once $\kappa > \sqrt{2}/2$, and
a constant calibrated on hundred-candidate shortlists would determine whether small candidate sets are demoted
at all. We therefore apply the relative cutoff only when the shortlist holds at least four candidates and use the
absolute rule otherwise.

The condition depends on the size of the shortlist and not on the evidence it contains. A condition on the
evidence would also disable the cutoff on a flat hundred-candidate shortlist, which is the case
Proposition~\ref{prop:flat} covers. As a consequence, the Boolean suite falls entirely under the absolute rule,
because every BoolQuestions candidate set holds two or three documents; it therefore measures the soft penalty alone, with no hard demotion. The six exclusion collections and the three no-harm suites use top-$100$
shortlists, where the condition never applies and the results are identical with or without it.

\paragraph{Learning rates.} The learning rate is selected in the same way for every trainable system, including EXCISE and the external baselines: by five-fold cross-validation with folds grouped by positive document id, so that no positive document is shared between folds. Ties are resolved in favour of the smaller rate, and the selected rate is then used to retrain
on all $1{,}900$ triples. The selected rates differ widely across architectures, so a single shared rate would
have favoured some architectures over others.

\FloatBarrier
\subsection{Design Alternatives}
\label{app:design}

Beyond the constants, three design choices in the demotion stage could be questioned. This section tests each
of them on Reason-ModernColBERT with everything else held fixed (Table~\ref{tab:design}).

The first is pooling: the evidence score averages over the tokens of the excluded topic, whereas taking the
strongest match might seem more natural for deciding whether a document covers a topic. The second is the
number of demoted candidates, which adapts to the evidence instead of being fixed. The third is the scoring
query: since the inversion arises from the contribution of the excluded tokens to MaxSim, relevance could be
scored without those tokens.

\begin{table*}[t]\centering
\caption{\textbf{Design alternatives, each swapped into the deployed system one at a time.} Exclusion success@10 per collection; \textbf{bold} marks the column maximum. $^{\dagger}$This row switches the adaptive rule off, so it is a different operator rather than a tightening of the deployed one. Reason-ModernColBERT, the six test collections.}
\label{tab:design}
\resizebox{\textwidth}{!}{%
\begin{tabular}{@{}llcccccc@{}}
\toprule
\textbf{Variant} & \textbf{change from deployed} & ExcluIR & FiQA & TREC-C & ESGen & EDGAR & EUR-Lex \\
\midrule
\rowcolor{gray!12}\multicolumn{8}{c}{\textit{deployed operator}} \\
\oursrow \textbf{\excise{}} & -- & 0.8502 & 0.7548 & 0.8455 & \textbf{0.6553} & \textbf{0.6771} & \textbf{0.7449} \\
\midrule
\rowcolor{gray!12}\multicolumn{8}{c}{\textit{how the excluded topic's evidence is pooled}} \\
\;max & strongest $Z$-token match & 0.8159 & 0.6875 & 0.7580 & 0.4806 & 0.6062 & 0.6582 \\
\;top-$m$ & mean of the top two & \textbf{0.8610} & 0.7212 & 0.8105 & 0.6019 & 0.6516 & 0.7092 \\
\rowcolor{gray!12}\multicolumn{8}{c}{\textit{what relevance is scored on}} \\
\;mask-$Z$ & score on the wanted tokens only & 0.7906 & 0.7548 & 0.8076 & 0.6165 & 0.6686 & 0.7092 \\
\rowcolor{gray!12}\multicolumn{8}{c}{\textit{how many candidates are demoted}} \\
\;cap 1 & adaptive rule, capped at one & 0.8556 & \textbf{0.7596} & 0.8455 & 0.6456 & \textbf{0.6771} & \textbf{0.7449} \\
\;fixed 1 & demote exactly one & 0.8556 & \textbf{0.7596} & 0.8455 & 0.6456 & \textbf{0.6771} & \textbf{0.7449} \\
\;fixed 2$^{\dagger}$ & demote exactly two & 0.8520 & 0.7548 & 0.8455 & 0.6505 & 0.6742 & \textbf{0.7449} \\
\rowcolor{gray!12}\multicolumn{8}{c}{\textit{hard-demotion fraction} $\phi$} \\
\;$\phi = 0.80$ &  & 0.8520 & 0.7548 & \textbf{0.8484} & \textbf{0.6553} & 0.6742 & \textbf{0.7449} \\
\;$\phi = 0.85$ &  & 0.8520 & 0.7548 & \textbf{0.8484} & \textbf{0.6553} & 0.6742 & \textbf{0.7449} \\
\;$\phi = 0.90$ &  & 0.8520 & \textbf{0.7596} & \textbf{0.8484} & \textbf{0.6553} & 0.6714 & \textbf{0.7449} \\
\rowcolor{gray!12}\multicolumn{8}{c}{\textit{combination}} \\
\;mask-$Z$ $+$ top-$m$ &  & 0.7960 & 0.7308 & 0.7843 & 0.6019 & 0.6346 & 0.6684 \\
\bottomrule
\end{tabular}}
\end{table*}

\paragraph{Pooling.} Averaging acts as a regulariser. Replacing the mean with top-$m$ pooling lowers
success@10 on five of the six collections, most on ESGenius, which falls from $0.6553$ to $0.6019$. A pure
maximum makes the mechanism clear: it scores $0.8159$ on ExcluIR and $0.4806$ on ESGenius, a decrease of
$0.175$ on the latter.

Under a maximum, a single excluded token that matches strongly is sufficient to demote a document. Gold
documents frequently share vocabulary with the excluded topic, so they are demoted together with the
negatives. Averaging prevents the penalty from responding to such coincidental matches.

\paragraph{The hard-demotion count.} A compound query names two excluded topics and has two confusable
negatives, so demoting more candidates could be expected to help. It could equally be expected to harm,
because on a single-exclusion query the second demoted candidate may be a wanted document.

Neither effect is observed. Every rule from a cap of one to $\phi = 0.90$ scores between $0.8502$ and $0.8556$
on ExcluIR, against a binomial standard error of about $0.015$ at $n = 554$. A larger cap does no measurable harm: a candidate is eligible for hard demotion only if it already stands out from the evidence distribution of the shortlist, so the additional candidates it demotes are rarely the gold document.

We adopt the adaptive rule because it responds to the evidence rather than to an assumed number of exclusions.
Under this rule a fixed demotion count is \textbf{not a free parameter}: the set of demoted candidates is
determined by the evidence fraction and truncated by the cap. Fixing the count at one is equivalent to capping
the adaptive rule at one, since both select the candidate with the highest evidence. The cap was fixed before
the sweep, and the differences observed are too small to justify changing it.

\paragraph{Scoring on the wanted query alone.} Removing the excluded tokens from the scoring query lowers
success@10 on five of the six collections, for example from $0.8502$ to $0.7906$ on ExcluIR and from $0.6553$
to $0.6165$ on ESGenius. This comparison has one caveat: the exclusion adapter was \emph{trained} on full
queries, so masking at inference is outside its training distribution.

\paragraph{Shortlist size.} The three alternatives above change what the demotion does, whereas the shortlist
depth changes only its cost. Table~\ref{tab:topk} varies $k$. Halving the shortlist lowers success@10 on five
of the six collections, by up to $0.034$ on ESGenius, and raises it only on ExcluIR, by $0.007$. The sweep
therefore supports the deployed $k = 100$, which was fixed beforehand; the query-time figures show that the
additional re-embedding cost is small.

\begin{table}[t]\centering
\caption{\textbf{Exclusion success@10 as the re-embedded shortlist shrinks}, on Reason-ModernColBERT. Halving it lowers success@10 on five of the six collections and raises it only on ExcluIR, which supports the deployed $k = 100$; Appendix~\ref{app:design} discusses the sweep.}
\label{tab:topk}
\resizebox{\columnwidth}{!}{%
\begin{tabular}{@{}lcccccc@{}}
\toprule
\textbf{Shortlist} & ExcluIR & FiQA & TREC-C & ESGen & EDGAR & EUR-Lex \\
\midrule
$k = 25$ & 0.8502 & 0.7019 & 0.8017 & 0.6165 & 0.6856 & 0.6990 \\
$k = 50$ & 0.8574 & 0.7404 & 0.8251 & 0.6214 & 0.6714 & 0.7143 \\
$k = 75$ & 0.8592 & 0.7548 & 0.8426 & 0.6505 & 0.6912 & 0.7143 \\
$k = 100$ \textit{(deployed)} & 0.8502 & 0.7548 & 0.8455 & 0.6553 & 0.6771 & 0.7449 \\
\bottomrule
\end{tabular}}
\end{table}

\paragraph{Why demoting harder does not help.} Most of the remaining errors are gold documents that leave the
top ten, not excluded documents that remain in it: on ExcluIR the gold is in the top ten for $0.877$ of queries
and leak is only $0.021$ (Table~\ref{tab:tm_excluir}). Demoting more aggressively does not help, because gold
documents frequently share vocabulary with the excluded topic, so a rule that demotes more also demotes gold documents
(Table~\ref{tab:design}). The deployed configuration is therefore a point on a trade-off rather than an optimum.

\paragraph{Synonyms and paraphrase.} The evidence score compares contextualised token embeddings, so it does
not require the excluded topic to be named verbatim. On the $683$ ExcluIR and FiQA queries whose excluded
document names the topic verbatim, \excise{} and the sparse BM25-$Z$ rule of Table~\ref{tab:ninefix} score
$0.837$ and $0.810$ with the detector's topic. On the remaining $79$ queries, \excise{} still reaches $0.709$, whereas the sparse rule
falls to $0.506$.

\subsection{Two Worked Examples}
\label{app:example}

The operator is easier to follow on a single query than in aggregate. The two test queries below are among
those on which the frozen index, every fine-tuned ColBERT baseline and the strongest reranker all fail. The
first is cue-free and the second is compound.

\begin{tcolorbox}[colback=gray!5, colframe=gray!75, title={A cue-free query (T2), end to end}, fonttitle=\bfseries\small, breakable]
\small
\textbf{Query:} ``Beyond using retirement money to buy a home, how should home ownership affect retirement
savings calculations?''\\
\textbf{What is being asked:} how a home should count in retirement savings. No word is negated; only
\emph{beyond} signals that advice on using retirement money to buy a home is not wanted.\\
\textbf{Frozen ranking:} the excluded answer, which warns against using retirement money for a home, is ranked
first, and the gold is third.\\
\textbf{Other systems:} the fine-tuned baselines keep the excluded answer in the top ten. The 568M reranker
removes it, but also pushes the gold out of the top ten.\\
\textbf{Detector output:} ``using retirement money to buy a home''; the switch fires.\\
\textbf{After \excise{}:} the gold rises to rank 2, and the excluded answer falls to rank $98$.
\end{tcolorbox}

\noindent The compound tier is the hardest, because a single query excludes two topics, each with its own
confusable document.

\begin{tcolorbox}[colback=gray!5, colframe=gray!75, title={A compound query (T3), end to end}, fonttitle=\bfseries\small, breakable]
\small
\textbf{Query:} ``With Mass Observation already covered, I want British documentary series about art and
culture, but not Kenneth Clark. Please suggest well-matched alternatives.''\\
\textbf{What is being asked:} British documentary series about art and culture, excluding two topics: Mass
Observation, implicitly through \emph{already covered}, and Kenneth Clark, explicitly through \emph{but not}.\\
\textbf{Frozen ranking:} the pages on Humphrey Jennings (a founder of Mass Observation) and Kenneth Clark are
ranked first and second; the gold, \emph{The Victorians}, is third.\\
\textbf{Other systems:} each fine-tuned baseline keeps at least one excluded page in the top ten, and the 568M
reranker does not return the gold in its top 100.\\
\textbf{Detector output:} ``Mass Observation Kenneth Clark''; the switch fires and recovers both topics.\\
\textbf{After \excise{}:} the gold is ranked first, and the two excluded pages fall to ranks $98$ and $99$.
\end{tcolorbox}

\section{The \xbench{} Benchmark}
\label{app:partC}

\subsection{Construction and Prompts}
\label{app:bench}

This section describes the construction of \xbench{} and gives the prompts used. Table~\ref{tab:data} lists the resulting collections.

Each corpus is divided into indexed documents at the unit that its source provides. ExcluIR, FiQA and TREC-COVID already consist of short documents, which are kept whole. ESGenius is a collection of report PDFs that have no structure below the page level, and its questions cite pages, so we divide it by page (a median of $463$ words). EDGAR and EUR-Lex documents are much longer, so we divide them at their existing boundaries: EDGAR by
section ($227$ words) and EUR-Lex by article ($137$ words).

\begin{table}[t]\centering
\caption{\textbf{\xbench{} evaluation suites.} Exclusion collections with their corpus size, query count and tier composition; the no-harm and Boolean control suites use existing human relevance judgements only. ExcluIR and FiQA are in domain; the other four are held out entirely and are zero-shot for every system compared here.}
\label{tab:data}
\resizebox{\columnwidth}{!}{%
\begin{tabular}{@{}llrrrrr@{}}
\toprule
\textbf{Role} & \textbf{Dataset} & \textbf{Corpus} & \textbf{T1} & \textbf{T2} & \textbf{T3} & \textbf{Total} \\
\midrule
\rowcolor{gray!12}\multicolumn{7}{c}{\textit{Exclusion: in-domain (disjoint train/test)}} \\
In-dom. & ExcluIR & 90{,}406 & 199 & 172 & 183 & \textbf{554} \\
In-dom. & FiQA & 57{,}638 & 77 & 62 & 69 & \textbf{208} \\
\midrule
\rowcolor{gray!12}\multicolumn{7}{c}{\textit{Exclusion: out-of-domain (zero-shot)}} \\
OOD & TREC-COVID & 171{,}332 & 137 & 103 & 103 & \textbf{343} \\
OOD & ESGenius & 995 & 82 & 62 & 62 & \textbf{206} \\
OOD & EDGAR & 676{,}925 & 119 & 111 & 123 & \textbf{353} \\
OOD & EUR-Lex & 322{,}938 & 79 & 59 & 58 & \textbf{196} \\
\midrule
\rowcolor{gray!12}\multicolumn{7}{c}{\textit{Control suites (existing human judgements)}} \\
No-harm & SciFact / NFCorpus / ArguAna & \multicolumn{4}{c}{300 / 323 / 1{,}401 queries} & \textbf{2{,}024} \\
Boolean & BoolQuestions (AND/OR/NOT)   & \multicolumn{4}{c}{pairwise-scorable pairs} & \textbf{834} \\
\bottomrule
\end{tabular}}
\end{table}

\paragraph{Mining confusable pairs.} The difficulty of the benchmark comes from its negatives. For each in-domain corpus, we embed every document with the frozen backbone and mine up to two \emph{confusable negatives} per anchor document within a similarity band, which is widened for heterogeneous corpora. A confusable negative shares the wanted topic with the gold but also covers an excluded topic: it is the document that a retriever should rank highly on grounds of relevance and must nevertheless suppress. The out-of-domain sets are mined in the same way and are never used for training.

\paragraph{The nine gates.} A generated record is admitted only if it passes all nine checks below. They are applied to the final query, once generation is complete.

\begin{enumerate}[nosep,leftmargin=*]
\item \textbf{Round-trip} (retrieval). The full query must retrieve the gold: a mean-pool shortlist is followed by a MaxSim re-ranking, and the gold must appear in the top $100$.
\item \textbf{Difficulty} (retrieval). Every confusable negative must be competitive with the gold under \emph{both}
BM25 and ColBERT. The negative and the gold are scored on the shared wanted topic $X$ rather than on the query,
so that confusability remains a property of the document pair and is therefore independent of the tier.
\item \textbf{$Z \in n$} (semantic). The judge must confirm that the excluded topic is present in the confusable negative $n$.
\item \textbf{$Z \notin g$} (semantic). The excluded topic must be absent from the gold $g$, checked both lexically and
by the judge.
\item \textbf{Distinctness} (structural). The gold and every negative must not be near-duplicates of one another.
\item \textbf{Multi-$Z$ coverage} (structural). In a compound query, each excluded topic must appear in \emph{its own} negative; two topics are never combined in one negative.
\item \textbf{Leakage} (structural). A MinHash near-duplicate check, at a Jaccard threshold of $0.9$, against
MS MARCO~\citep{nguyen2016msmarco} and the backbone corpora.
\item \textbf{Paraphrase safety} (semantic). On the cue-free and compound tiers a verbatim string match is not a
valid test, so the judge verifies again that the paraphrased topic is present in the negative and absent from the gold.
\item \textbf{Query-side judge} (semantic). For each excluded topic, the judge reads the query alone and must confirm
that the query still rules that topic out.
\end{enumerate}

\noindent Gates 3, 4 and 8 constitute the document-side semantic judge, which is applied once to each pair of excluded topic and negative. Gate 9 is the query-side judge, which detects a fluent rewrite that silently drops an exclusion.

Two of the nine gates, round-trip and difficulty, require separate comment, because both introduce an asymmetry that affects how the external comparison of Appendix~\ref{app:strong} should be interpreted. The \emph{round-trip} gate retains only queries whose gold can still be retrieved by a ColBERT pass; it is applied at construction, before any method
development. Its purpose is to make a failure of the frozen index attributable to the exclusion and not to an unreachable gold. The gate is, however, defined with respect to ColBERT, which the external retrievers do not share, because they retrieve their own shortlists.

The \emph{difficulty} gate has the same property. Negatives must be competitive under both BM25 and ColBERT, so difficulty is defined in the embedding space of ColBERT and the lexicon of BM25. Both gates are therefore shared by
every ColBERT-based system and by none of the others.

\paragraph{Generation.} The records filtered by these gates are produced in stages, and the procedure is template-free: stock phrases are banned, and the wanted and excluded topics are grounded in the mined documents and not invented. \textbf{Stage~A} produces the explicit
T1 query directly from a confusable pair. \textbf{Stage~B} rewrites it into the other tiers: T2 makes the exclusion implicit, and T3 adds a second excluded topic. A long form places the exclusions in the middle of the text instead of in a trailing clause. The queries are written by GPT models, while every gate judgement is made by
Llama-3.3-70B~\citep{grattafiori2024llama3}, so the model that writes a query is never the model that admits
it. All calls return strict JSON. The three tier prompts follow in abbreviated form.

\begin{tcolorbox}[colback=gray!5, colframe=gray!75, title=Stage A: T1 (Explicit) Generation, fonttitle=\bfseries\small, breakable]
\small
Given \textsc{doc\_p} (wanted) and \textsc{doc\_n} (must be excluded) from a confusable pair: name $X$ = the topic \emph{both} documents share, and $Z$ = the topic \emph{only} \textsc{doc\_n} covers (the reason it must be excluded). Write a natural query a real user would type that \textsc{doc\_p} answers and \textsc{doc\_n} violates \emph{because} it covers $Z$. Hard requirements: \textbf{(1)}~state the exclusion of $Z$ \textbf{explicitly}, with a negation cue (not / without / except / other than); \textbf{(2)}~contain $Z$ \textbf{copied verbatim}, never paraphrase or generalize it (if $Z=$``1982 film'', write ``\ldots not the 1982 film'', never ``not from the 80s''); \textbf{(3)}~do not exclude $X$-content. Vary the surrounding phrasing; avoid stock phrases. Return \texttt{\{"X":..,"Z":..,"query":..\}}.
\end{tcolorbox}

\begin{tcolorbox}[colback=gray!5, colframe=gray!75, title=Stage B: T2 (Implicit) Rewrite, fonttitle=\bfseries\small, breakable]
\small
Given the T1 query and its $X$, $Z$: rewrite so the exclusion of $Z$ is \textbf{implicit}, signal that $Z$ is already-known or already-covered (e.g.\ ``already covered'', ``leaving aside the \ldots we know''). Use \textbf{no exclusion cue at all}: the words \emph{not / without / except / excluding / other than / apart from / aside from / besides / neither / nor / rather than / -free} are forbidden. Keep $X$ the focus; $Z$ may be paraphrased but must still denote the \emph{same entity}. Return \texttt{\{"query":..,"Z\_surface":..\}}.
\end{tcolorbox}

\begin{tcolorbox}[colback=gray!5, colframe=gray!75, title=Stage B: T3 (Compound) Rewrite, fonttitle=\bfseries\small, breakable]
\small
Given a base query, wanted topic $X$, and two excluded topics $Z_1$, $Z_2$: rewrite into \emph{one} natural query that wants $X$ and rules out \textbf{both}, \textbf{mixing} the two exclusion types, $Z_1$ \textbf{explicit} (a negation cue that names it) and $Z_2$ \textbf{implicit} (signaled as already-known/covered, no negation word, but still named). Embed both mid-sentence, connective-rich (and / but); no trailing ``excluding\ldots'' clause. Two few-shot examples fix the exact one-explicit/one-implicit split. Return \texttt{\{"query":..,"Z":[Z1\_surface, Z2\_surface]\}}. A labeled \emph{disjunctive} sub-type instead rules out both explicitly with a boolean cue (``neither $Z_1$ nor $Z_2$'' / ``not $Z_1$ or $Z_2$'').
\end{tcolorbox}

\paragraph{Compound construction.} A T3 query excludes two topics, and each topic has its \emph{own} confusable negative. We never place both excluded topics in a single negative, because the statement ``the topic is in the negative'' would then be ambiguous and the label could not be verified. The first excluded topic comes from the mined pair. A second confusable negative supplies the other, and its additional topic is named by a short GPT-4.1 call. Two compound shapes result: a \emph{disjunctive} form, which rules out both topics with a Boolean cue, and a \emph{mixed} form, which pairs one explicit exclusion with one implicit exclusion. Both shapes contain two excluded topics with two separate confusable negatives, which is what makes the tier compound; they differ only in whether the second exclusion carries a cue.

\paragraph{The query-side judge.} A Stage~B rewrite can silently \emph{drop} an exclusion, leaving fluent
text that no longer rules out what it should. Gate~9 therefore applies the query-side judge to the final query,
once per excluded topic and without access to any document. Its prompt is given below.

\begin{tcolorbox}[colback=gray!5, colframe=gray!75, title=QC: Query-Side Exclusion Judge, fonttitle=\bfseries\small, breakable]
\small
Given a query and an excluded topic $Z$, decide whether the query, read naturally, would lead a user to \textbf{reject} documents that are primarily about $Z$. Answer strictly \texttt{\{"excludes\_Z": true|false\}}. Do not read any document; judge only the query's intent.
\end{tcolorbox}

\subsection{Benchmark Validity}
\label{app:human}

Three measurements support the validity of the benchmark.

\paragraph{1. The tiers are lexically clean.} The tier prompts specify which negation cues each tier requires and which it forbids, so a tier label is reliable only if the generator complies with them. Because the distinction between explicit and cue-free queries is central to the paper's claims, we verify this compliance on every admitted query with a strict cue test (\emph{not}, \emph{without}, \emph{except}, \emph{excluding}, \emph{exclude},
\emph{other than}, \emph{apart from}, \emph{aside from}, \emph{besides}, \emph{neither}, \emph{nor},
\emph{non-}, \emph{omit}, \emph{rather than}, \emph{minus}, \emph{-free}). Every T1 query contains a cue ($693/693$), as does every T3 query ($598/598$), whereas no T2 query does ($0/569$).

\paragraph{2. An independent model agrees on the labels.} A Qwen2.5-72B judge~\citep{qwen2025qwen25}, from a different model family than the Llama-3.3-70B that runs the gates, re-judges a stratified sample of $288$
admitted queries on whether each excludes its stated topic. It agrees with the admitted label on $94.4\%$ of
the sample, and the disagreements are not evenly distributed: agreement is $100\%$ on T1 and T3, and $83.3\%$ on T2.
All disagreements fall in the cue-free tier, as expected, because deciding whether a cue-free query rules out its topic requires judgement. The T2 results should therefore be read in light of the difficulty of the tier.

\paragraph{3. There is no train/test leakage.} Under normalised exact
match, comparing the four training sets with the six test collections yields \textbf{zero} overlapping queries in all $24$ train--test pairs. Under the same match, the training positives and negatives share \textbf{zero} documents with the
$3{,}253$ gold and confusable-negative documents of the test sets.

Tracing the training positives to their source corpora confirms the out-of-domain grouping: \textbf{none of them comes from} ESGenius, EDGAR, EUR-Lex or TREC-COVID. These four collections are zero-shot for \excise{} and, because every baseline is fine-tuned on the same triples, for every baseline as well. The rehearsal set of Appendix~\ref{app:strong} is SciDocs, which is not among the test or control sets and was chosen for that reason.
\subsection{Human Validation}
\label{app:humanval}

These three measurements are all automatic; the fourth is not. To verify that the model-generated labels reflect
genuine human intent, two PhD students external to the project independently evaluated a sample of $150$
queries, stratified by tier and by collection. Each annotator was asked whether the query naturally requests the wanted topic ($X$) while ruling out the excluded topic ($Z$), according to the tier definitions.

Inter-annotator agreement was high (Cohen's $\kappa = 0.85$), and the annotators agreed with the assigned
benchmark labels on $91.3\%$ of the sample overall. Agreement was highest on the explicit T1 tier ($96.0\%$) and the compound T3 tier ($94.0\%$), and decreased to $84.0\%$ on the cue-free T2 tier. These results indicate that the benchmark labels are largely sound, although judging a cue-free exclusion is more subjective.

\subsection{The \xbench{} Query Gallery}
\label{app:gallery}

These judgements are easier to assess alongside the queries themselves. The gallery presents verbatim examples of
every query type, chosen to make the task concrete and to show what a learned operator must recover that a
lexical rule cannot. Each entry lists the wanted topic, the excluded topic
or topics, and a note on the source of difficulty. Table~\ref{tab:gallery-excl} covers T1 and T2 in short and
long forms, and T3 in both compound shapes.

\begin{table*}[t]\centering\footnotesize

\caption{\xbench{} exclusion gallery, drawn verbatim from the benchmark and selected by a fixed rule, one row per query type, so that all six collections are represented. T1 states the exclusion with a cue; T2 contains \emph{no} negation cue; T3 excludes two topics at once, either disjunctively or by combining one explicit and one implicit exclusion. T1 and T2 appear in a short and a long form, and T3 in a long form only; the long form places the exclusion in the middle of the sentence.}
\label{tab:gallery-excl}
\renewcommand{\arraystretch}{1.15}
\begin{tabular}{@{}p{1.85cm}p{13.55cm}@{}}
\toprule
\textbf{Type} & \textbf{Verbatim query, with wanted/excluded topics and the source of difficulty} \\
\midrule
\textbf{T1 explicit} \newline \emph{short} &
\emph{``Which capital investments tied to projected volume levels, excluding international operations, support existing business and anticipated growth?''} \newline
$X=$ capital investments;\; $Z=$ international operations (cue: \emph{excluding}). \; \emph{Why hard:} the excluded document also discusses the wanted topic, so it matches the wanted tokens, and the tokens naming the excluded topic raise its score further.\\
\addlinespace[2pt]
\textbf{T1 explicit} \newline \emph{long} &
\emph{``What requirements apply to conflict-of-interest disclosures when information is provided by means of a website, not the key information document, and how should those disclosures be presented?''} \newline
$X=$ information provided by means of a website;\; $Z=$ key information document (cue: \emph{not}). \; \emph{Why hard:} the exclusion is placed mid-sentence between two clauses of the wanted topic, and the two topics are instruments of the same regulation, so the documents that discuss either discuss both.\\
\midrule
\textbf{T2 cue-free} \newline \emph{short} &
\emph{``With case studies on Scope 3 greenhouse gas emissions already covered, what are the recent trends in corporate climate-related financial disclosures?''} \newline
$X=$ climate-related financial disclosures;\; $Z=$ case studies on Scope 3 emissions (\textbf{no negation cue}). \; \emph{Why hard:} the query contains no negation word; ``already covered'' alone carries the exclusion, so a lexical filter has no cue to match, and Scope 3 case studies are a standard subtopic of the wanted one.\\
\addlinespace[2pt]
\textbf{T2 cue-free} \newline \emph{long} &
\emph{``I'm researching 1970s American cinema for a class and have already covered the Blue Like Jazz film; which movie follows a businessman facing moral conflict in contemporary America and stars Jack Lemmon?''} \newline
$X=$ film;\; $Z=$ Blue Like Jazz (\textbf{no cue}; signalled by ``already covered''). \; \emph{Why hard:} the excluded title reads as \emph{background} rather than as a constraint, and the wanted film is given only by plot and cast, so no wanted token connects the query to its own gold.\\
\midrule
\textbf{T3 compound} \newline \emph{disjunctive} &
\emph{``How does dollar-cost averaging fit my plan to invest a fixed sum over ten months for an average entry price, when this is neither regular 401(k) contributions nor an employee stock purchase plan?''} \newline
$X=$ dollar-cost averaging;\; $Z_1=$ regular 401(k) contributions, $Z_2=$ employee stock purchase plan. \; \emph{Why hard:} both excluded topics are themselves forms of periodic investing, each with its own confusable document, and the query rules out their \emph{union}, so two documents must be demoted.\\
\addlinespace[2pt]
\textbf{T3 compound} \newline \emph{mixed} &
\emph{``During COVID-19, our clinic already has guidance on access to later abortion; what practical advice can help us maintain abortion as an essential health service, not frame the issue as a human rights imperative?''} \newline
$X=$ abortion as an essential health service;\; $Z_1=$ a human rights imperative (explicit: \emph{not});\; $Z_2=$ access to later abortion (implicit: \emph{already has guidance on}). \; \emph{Why hard:} the sentence contains two exclusions of \emph{different} kinds, so an operator that relies on the cue finds one of them and misses the other.\\
\addlinespace[2pt]
\textbf{T3 compound} \newline \emph{long, OOD} &
\emph{``With the ALLN-346 product candidate already covered, the focus is the market opportunity for enteric hyperoxaluria. How should that opportunity, but not other product candidates, be assessed?''} \newline
$X=$ enteric hyperoxaluria;\; $Z_1=$ other product candidates (explicit: \emph{but not});\; $Z_2=$ ALLN-346 (implicit: \emph{already covered}). \; \emph{Why hard:} one excluded topic is a named entity and the other an open class, so there is no single string to exclude, and ALLN-346 is the candidate developed for the wanted indication.\\
\bottomrule
\end{tabular}
\end{table*}

\section{Experimental Setup}
\label{app:partD}

\subsection{Configuration Details}
\label{app:setup}

This section reports the deployed configuration, its query-time cost and the recall of its first stage.
First-stage retrieval returns the top $k = 100$ candidates, and \excise{} re-embeds and re-ranks only these.
The deployed constants are $\theta = 0.70$ (firing threshold), $\tau = 0.35$ (evidence floor), $\kappa = 0.5$ (cutoff width), $\lambda = 0.97$ (penalty scale) and $\phi = 0.6$ with a cap of $3$ (hard demotion), together
with a minimum shortlist of $4$ candidates for the relative cutoff. Both adapters have rank 8 and are trained for
3 epochs; the learning rate of the exclusion adapter is $10^{-3}$ on all three backbones. All constants are
fixed on the training split (Appendix~\ref{app:sens}) and reused unchanged on every test set and every
backbone.

Document handling differs by backbone, because each backbone is kept within its own context length. GTE and ColBERTv2, which run with a 300-token document length, split documents into 200-word chunks, which are pooled by MaxSim, and cap documents at $1{,}000$ words. Reason uses 512-word chunks and a cap of $10{,}000$ words. Only $0.03\%$ of documents exceed $1{,}000$ words, so this difference cannot separate the backbones.

The chunk size is not equalised across backbones. A 512-word chunk corresponds to about $666$ tokens, so
applying Reason's chunk size to a 300-token encoder would truncate roughly $55\%$ of every chunk; these two backbones would then read less text while appearing to be treated equally. Chunk size is a
property of the encoder, and fairness is addressed through the document budget. The fine-tuned baselines are
re-indexed and queried with the adapted encoder, every evaluation uses the same shortlist depth, and every run
records its full operator configuration.

\paragraph{Query-time cost.} On one H100 with a warm index, the frozen first stage takes $72.3$\,ms. The
detector adds $12.7$\,ms to every query. On a query that fires, re-embedding and re-ranking the top $100$
candidates and applying the demotion add a further $15.1$\,ms, giving a total overhead of $27.8$\,ms, or about $0.4\times$ the frozen latency. The expected overhead of a workload with firing rate $\rho$ is
$12.7 + 15.1\rho$\,ms.

Most of the per-firing cost depends on one implementation detail. The adapter's encoding of a \emph{document}
depends only on the document and the adapter, because the query enters only at the MaxSim stage. A candidate
re-embedded for one query can therefore be reused for any later query whose shortlist contains it, and
shortlists overlap heavily. \excise{} keeps a bounded cache of re-embedded candidates, which is limited by a
memory budget and discarded whenever the index changes. The cache is not a second index: it holds only
documents that queries have retrieved, and it never requires a pass over the corpus. Document encoding accounts
for $87\%$ of the firing cost, so the figures above are steady-state values; with a cold cache, this $87\%$ is
paid on every query until shortlists begin to overlap. The cache is enabled in all experiments, so every
reported number is measured with it.

\paragraph{First-stage recall.} The operator can recover a gold document only if the frozen first stage places
it in the top $k = 100$. We measure recall@100 on the shortlists returned by the deployed system, which is a
different retrieval pass from the round-trip gate of Appendix~\ref{app:bench}. On Reason it is $1.000$ on
ExcluIR, $1.000$ on FiQA, $1.000$ on TREC-COVID, $0.985$ on ESGenius, $0.994$ on EDGAR and $1.000$ on EUR-Lex.
The shortlist is therefore nearly complete on every collection, and first-stage recall does not limit the
results.

\paragraph{The chunk unit and the demotion.} ColBERT scores each document by its best-matching chunk
(\S\ref{sec:setup}), and in \excise{} the demotion operates on the same unit. Only the best chunk of a
candidate is passed on, and re-embedding, evidence scoring and demotion read no other chunk. An excluded topic
that occurs elsewhere in the document therefore produces no evidence and no demotion. At Reason's $512$-word
chunk this situation is confined to ESGenius, where $44.4\%$ of confusable negatives exceed the chunk size and
$60.5\%$ of those contain a chunk in which the excluded topic does not appear; on the other five collections at
most $1.1\%$ of negatives are divided at all. The gold is retrieved as before, so the cost falls on suppression
and not on retrieval. Within each band of Table~\ref{tab:main} all systems read the same corpus at the same
chunk size, so those comparisons are unaffected. No baseline has a demotion stage that this condition could
affect, and the ESGenius margin on Reason, $0.291$ over the strongest baseline, was measured under it.

\subsection{External Baselines}
\label{app:strong}

This section describes how the external systems were configured and compared. Table~\ref{tab:strong} reports the comparison at our $10{,}000$-word budget, with each document read in $350$-word passages. Every external system was also run at its published 512-token limit, where the pooled success@10 of the strongest reranker is $0.8022$ instead of $0.7855$. We do not use that setting for the main comparison, because \excise{} always indexes the whole document; Table~\ref{tab:windowing} reports its effect per collection.

\begin{table*}[t]\centering
\caption{\textbf{Eight external systems at our $10{,}000$-word budget}, each off the shelf and fine-tuned (\emph{ft}). All fine-tuned rows were trained on the same $1{,}900$ triples under the same protocol. The collection columns give exclusion success@10; \textbf{Pooled} is success@10 over all $1{,}860$ queries and is the column tested by the paired bootstrap. Documents are windowed and max-pooled (Appendix~\ref{app:strong}). \textbf{No-harm} is mean nDCG@10 on SciFact, NFCorpus and ArguAna at the same budget, with the change from the off-the-shelf model in parentheses. The systems start from different levels, so this change, not the absolute level, measures the cost to ordinary retrieval.}
\label{tab:strong}
\resizebox{\textwidth}{!}{%
\begin{tabular}{@{}lr cc cccc c c@{}}
\toprule
\textbf{System} & \textbf{Size} & \multicolumn{2}{c}{\textbf{In-domain}} & \multicolumn{4}{c}{\textbf{Out-of-domain (zero-shot)}} & \textbf{Pooled} & \textbf{No-harm} \\
\cmidrule(lr){3-4}\cmidrule(lr){5-8}\cmidrule(lr){9-9}\cmidrule(lr){10-10}
 & & ExcluIR & FiQA & TREC-C & ESGen & EDGAR & EUR-Lex & \textit{n}=1{,}860 & nDCG@10 \\
\midrule
\rowcolor{gray!12}\multicolumn{10}{c}{\textit{Off the shelf, no re-index}}\\
BGE-large & 335M & 0.1877 & 0.2548 & 0.3120 & 0.1165 & 0.2691 & 0.3622 & 0.2441 & 0.5298 \\
E5-large & 335M & 0.1787 & 0.3558 & 0.3440 & 0.1650 & 0.3173 & 0.3571 & 0.2726 & 0.4698 \\
gte-Qwen2-1.5B & 1.5B & 0.1552 & 0.2548 & 0.2595 & 0.0922 & 0.2606 & 0.3265 & 0.2166 & 0.5512 \\
Promptriever (mistral-7B, instr.) & 7B & 0.3484 & 0.3029 & 0.4373 & 0.1214 & 0.3768 & 0.4439 & 0.3500 & 0.4542 \\
MS-MARCO MiniLM & 22M & 0.1931 & 0.2596 & 0.3848 & 0.0631 & 0.2521 & 0.3418 & 0.2484 & 0.4556 \\
bge-reranker-base & 278M & 0.1282 & 0.1875 & 0.2653 & 0.0825 & 0.2890 & 0.3929 & 0.2135 & 0.3947 \\
bge-reranker-large & 560M & 0.1408 & 0.2019 & 0.2915 & 0.0583 & 0.2776 & 0.3776 & 0.2172 & 0.4260 \\
bge-reranker-v2-m3 & 568M & 0.1336 & 0.1923 & 0.2391 & 0.0825 & 0.2861 & 0.3571 & 0.2065 & 0.4542 \\
\addlinespace[2.5pt]
\rowcolor{gray!12}\multicolumn{10}{c}{\textit{Fine-tuned retrievers, full re-index required}}\\
BGE-large (ft) & 335M & 0.6588 & 0.5144 & 0.6968 & 0.3010 & 0.4476 & 0.5561 & 0.5591 & 0.4980\,($-$.032) \\
E5-large (ft) & 335M & 0.6354 & 0.4856 & 0.6210 & 0.2427 & 0.5099 & 0.5663 & 0.5414 & 0.4364\,($-$.033) \\
gte-Qwen2-1.5B (ft) & 1.5B & 0.7148 & 0.5337 & 0.5977 & 0.2524 & 0.4193 & 0.5357 & 0.5468 & 0.3442\,($\mathbf{-}$\textbf{.207}) \\
Promptriever (ft) & 7B & 0.7870 & 0.6298 & 0.7464 & 0.2767 & 0.4986 & 0.6633 & 0.6376 & 0.4876\,($+$.033) \\
\addlinespace[2.5pt]
\rowcolor{gray!12}\multicolumn{10}{c}{\textit{Fine-tuned cross-encoder rerankers, BGE-large first stage, top-100 re-scored, no re-index}}\\
MS-MARCO MiniLM (ft) & 22M & 0.6733 & 0.5577 & 0.5627 & 0.5049 & 0.4589 & 0.6224 & 0.5753 & 0.3603\,($-$.095) \\
bge-reranker-base (ft) & 278M & 0.8231 & 0.6635 & 0.6939 & 0.5194 & 0.6091 & 0.7041 & 0.6946 & 0.3520\,($-$.043) \\
bge-reranker-large (ft) & 560M & 0.8087 & 0.6538 & 0.6706 & 0.5874 & 0.5864 & 0.6786 & 0.6855 & 0.3608\,($-$.065) \\
bge-reranker-v2-m3 (ft) & 568M & 0.8845 & 0.7548 & 0.8601 & 0.6699 & 0.6544 & 0.7653 & 0.7855 & 0.3738\,($-$.080) \\
\addlinespace[2.5pt]
\rowcolor{gray!12}\multicolumn{10}{c}{\textit{ColBERT (Reason), frozen index, no re-index}}\\
Frozen ColBERT & 149M & 0.1354 & 0.1827 & 0.1866 & 0.1699 & 0.3031 & 0.2806 & 0.2011 & 0.5126 \\
\oursrow \textbf{\excise{}} (ours) & 149M\,$+$\,1.5M & 0.8502 & 0.7548 & 0.8455 & 0.6553 & 0.6771 & 0.7449 & 0.7731 & 0.5072\,($-$.005) \\
\bottomrule
\end{tabular}}
\end{table*}

\paragraph{Models.} The dense retrievers are \texttt{BAAI/\allowbreak bge-\allowbreak large-\allowbreak en-\allowbreak v1.5}, \texttt{intfloat/\allowbreak e5-\allowbreak large-\allowbreak v2} and
\texttt{Alibaba-NLP/\allowbreak gte-\allowbreak Qwen2-\allowbreak 1.5B-\allowbreak instruct} (last-token pooling, its native instruction prefix), each queried
against its own single-vector index built over the same corpus. The cross-encoders are
\texttt{cross-encoder/\allowbreak ms-\allowbreak marco-\allowbreak MiniLM-\allowbreak L-6-v2} (22M), \texttt{BAAI/\allowbreak bge-\allowbreak reranker-\allowbreak base} (278M),
\texttt{BAAI/\allowbreak bge-\allowbreak reranker-\allowbreak large} (560M) and \texttt{BAAI/\allowbreak bge-\allowbreak reranker-\allowbreak v2-m3} (568M), each re-scoring a
BGE-large top-100. The instruction-following
retriever is Promptriever~\citep{weller2024promptriever}, specifically
\texttt{samaya-ai/\allowbreak promptriever-\allowbreak mistral-v0.1-7b-v1}; the family has several variants, which are not
interchangeable. Promptriever receives the exclusion as a natural-language instruction appended to the
query, as its model card prescribes. All systems see the same test queries and are scored with the same metric.

\paragraph{Shared supervision and matched tuning.} Each external baseline is fine-tuned on the same $1{,}900$
retrieval triples as every other trainable system, including \excise{}'s adapter. The adapter is additionally
trained on $727$ demotion triples (\S\ref{sec:bench}). These supervise a subtracted evidence term that no
baseline has; used as ordinary positives, they would instruct a baseline to rank the excluded document highly.
The retrievers are trained with \textsc{MultipleNegativesRankingLoss} for 3 epochs with each model's native
query and document prefixes, after which the corpus is re-encoded and re-indexed with the fine-tuned weights.
This setup serves as a placement control: the retrieval supervision that \excise{} confines to a small
query-time adapter is given to the full encoder instead, so a difference in the table reflects \emph{where}
the supervision acts. The training set is not restricted to exclusion: $445$ of the $1{,}900$ triples are questions without an exclusion, so every system also sees ordinary retrieval. The training setup does not include rehearsal on a corpus outside the evaluation; we add and measure this below. No system is tuned against the no-harm results.

\paragraph{Windowing and its control.} A fair comparison also depends on how much of each document a system
reads. Documents are split into 350-word passages, the largest that fit a 512-token encoder without
truncation. Scores are max-pooled over passages, and documents are capped at $10{,}000$ words to match our own
setting.

ExcluIR serves as a control. Its documents are short, so windowing leaves nearly all of them intact: $90{,}406$
documents yield $90{,}420$ passages, so that fourteen documents are split. A change on ExcluIR would therefore
indicate a defect in the windowing rather than an effect of it. Of the sixteen external rows measured at both
budgets, thirteen reproduce their truncated ExcluIR success@10 exactly and twelve reproduce their leak. The
three that change on success@10 ($0.1570 \to 0.1552$, $0.3466 \to 0.3484$ and $0.7798 \to 0.7870$) differ by
one to four queries. The four that change on leak belong to the two last-token-pooled models, off the shelf and fine-tuned, where a maximum over two passages can change a single rank. All cross-encoder rows and the BGE and E5 rows
agree to four decimals on both metrics.

\paragraph{Where windowing has an effect.} The effect is confined to one collection, where it applies to every
system. ESGenius is the only collection whose documents are routinely divided into more than one passage.
There, windowing raises the leak of all eight fine-tuned external systems: by $0.129$, $0.131$ and $0.134$ for
the three bge rerankers, by $0.061$ for MiniLM, and by $0.004$ to $0.206$ for the four single-vector
retrievers. Success@10 falls for each of them, by $0.034$ to $0.131$. On the other five collections, the
largest change in leak for any system is $0.014$.

The effect is caused by aggregation over passages, not by truncation. On ESGenius the excluded topic occurs at
a median position of word $18$ in the confusable negative, and only $1.1\%$ of its occurrences lie beyond the
512-token limit, so a truncated model still reads it. Windowing divides the document and scores it by its best
passage, and $71.1\%$ of the ESGenius negatives that are divided contain a passage in which the excluded topic
does not appear. Retrieval is unaffected: the 568M reranker's hit@10 changes from $0.845$ to $0.850$, while
its leak rises from $0.034$ to $0.163$. Scoring the same documents by their worst passage reduces that leak to
$0.019$, which isolates the mechanism. The effect does not depend on architecture: the largest change is
Promptriever's $+0.206$, a single-vector retriever, and the smallest is E5's $+0.004$. It depends on how often
documents are divided, which explains why the leak in the other five rows of Table~\ref{tab:windowing} changes by at most $0.007$: on those collections at most $4.3\%$ of confusable negatives produce more than one passage. \excise{}
is also affected: at its $512$-word chunk its leak on ESGenius is $0.153$, compared with the reranker's
$0.163$, so both are exposed to the same effect at different window widths.

\begin{table}[t]\centering
\caption{\textbf{Effect of windowing on cross-encoder leak.} Rows are ordered by the share of corpus text that a 512-token limit discards. $\Delta$\,leak is the change in leak of the 568M and 278M rerankers when they read the whole document in 350-word passages instead of only its first 512 tokens. The next column gives the success@10 of the 568M reranker in both settings. \textbf{Bold} marks its two changes on ESGenius, the only collection whose documents are routinely divided into several passages. \excise{} indexes the whole document in both settings, so a single value is given for it.}
\label{tab:windowing}
\resizebox{\columnwidth}{!}{%
\begin{tabular}{@{}lccccc@{}}
\toprule
 & \textbf{text} & \multicolumn{2}{c}{$\boldsymbol{\Delta}$ \textbf{leak}} & \textbf{568M succ@10} & \textbf{\excise{}} \\
\cmidrule(lr){3-4}
\textbf{Collection} & \textbf{cut} & 568M & 278M & 512 tok. $\to$ windowed & succ@10 \\
\midrule
ESGenius & 25.6\% & $\mathbf{+0.129}$ & $+$0.134 & $\mathbf{0.8010 \to 0.6699}$ & 0.6553 \\
FiQA & 5.9\% & $+$0.002 & $+$0.002 & $0.7644 \to 0.7548$ & 0.7548 \\
TREC-COVID & 1.7\% & $+$0.003 & $-$0.007 & $0.8630 \to 0.8601$ & 0.8455 \\
EUR-Lex & 1.2\% & $+$0.000 & $+$0.000 & $0.7653 \to 0.7653$ & 0.7449 \\
EDGAR & 0.9\% & $+$0.000 & $+$0.000 & $0.6572 \to 0.6544$ & 0.6771 \\
ExcluIR & 0.0\% & $+$0.000 & $+$0.000 & $0.8845 \to 0.8845$ & 0.8502 \\
\bottomrule
\end{tabular}}
\end{table}

\paragraph{Re-ranking the shortlist of \excise{}.} Windowing equalises how much of a document each system
reads, but each system still retrieves its own candidates, so the comparison mixes retrieval with exclusion.
To separate the two, every fine-tuned external system is given \excise{}'s top-100 for each query and re-ranks
it. The excluded document is in that shortlist for $96\%$ of queries, so a system inherits the retrieval but
not the exclusion. Seven of the eight systems score below \excise{} (Table~\ref{tab:sameshort}). The
exception is bge-reranker-v2-m3, at $0.8344$ against $0.7731$. Model size does not explain this result:
bge-reranker-large, which differs by 8M parameters, scores below \excise{}, as do the 7B
instruction-following retriever at $0.6263$ and the 1.5B dense encoder at $0.5468$. The advantage also has a
cost: the reranker requires $100$ joint forward passes for every query, none of which can be reused for the
next query. The four single-vector systems show what a shared shortlist does not provide: given the same
candidates, their leak is $0.24$ to $0.34$, against $0.03$ to $0.05$ for the rerankers, because a shortlist
supplies candidates but no means of ruling one out.

\begin{table*}[t]\centering
\caption{\textbf{Every fine-tuned external system re-ranking the top-100 of \excise{}.} Each system receives the candidates that \excise{} retrieves for a query and re-orders them, so retrieval is shared and only exclusion is compared. The collection columns give exclusion success@10, and \textbf{Pooled} is over all $1{,}860$ queries. \textbf{Leak} is measured on the top ten of the shared candidates. The first row is \excise{}'s own ranking of its shortlist. Four systems leak less than \excise{} on the same candidates, yet three of them score lower overall, because a ranking that removes excluded documents also removes gold documents.}
\label{tab:sameshort}
\resizebox{\textwidth}{!}{%
\begin{tabular}{@{}lr cc cccc c c@{}}
\toprule
\textbf{System} & \textbf{Size} & \multicolumn{2}{c}{\textbf{In-domain}} & \multicolumn{4}{c}{\textbf{Out-of-domain (zero-shot)}} & \textbf{Pooled} & \textbf{Leak} \\
\cmidrule(lr){3-4}\cmidrule(lr){5-8}\cmidrule(lr){9-9}\cmidrule(lr){10-10}
 & & ExcluIR & FiQA & TREC-C & ESGen & EDGAR & EUR-Lex & \textit{n}=1{,}860 & $\downarrow$ \\
\midrule
\oursrow \textbf{\excise{}} (its own shortlist) & 149M\,$+$\,1.5M & 0.8502 & 0.7548 & 0.8455 & 0.6553 & 0.6771 & 0.7449 & 0.7731 & 0.075 \\
\midrule
bge-reranker-v2-m3 & 568M & 0.9206 & 0.7788 & 0.8630 & 0.6699 & 0.8074 & 0.8214 & 0.8344 & 0.025 \\
bge-reranker-large & 560M & 0.8321 & 0.6971 & 0.6939 & 0.5874 & 0.7025 & 0.7398 & 0.7301 & 0.028 \\
bge-reranker-base & 278M & 0.8357 & 0.6923 & 0.6822 & 0.5340 & 0.7422 & 0.7296 & 0.7290 & 0.035 \\
Promptriever (mistral-7B, instr.) & 7B & 0.7491 & 0.5769 & 0.6880 & 0.2670 & 0.5892 & 0.6684 & 0.6263 & 0.238 \\
MS-MARCO MiniLM & 22M & 0.6480 & 0.6010 & 0.5948 & 0.4757 & 0.5722 & 0.6378 & 0.5984 & 0.050 \\
gte-Qwen2-1.5B & 1.5B & 0.7058 & 0.5096 & 0.5918 & 0.2524 & 0.4533 & 0.5357 & 0.5468 & 0.242 \\
BGE-large & 335M & 0.6245 & 0.4856 & 0.6210 & 0.2864 & 0.4561 & 0.5561 & 0.5317 & 0.322 \\
E5-large & 335M & 0.5903 & 0.4808 & 0.5743 & 0.2379 & 0.4844 & 0.5765 & 0.5145 & 0.338 \\
\bottomrule
\end{tabular}}
\end{table*}

\paragraph{No-harm for the external systems.} The second axis is ordinary retrieval, measured as nDCG@10 on
SciFact, NFCorpus and ArguAna (BEIR test qrels, graded, scored with \texttt{pytrec\_eval}) and computed for
each system against its own embeddings. None of the three suites is free of truncation: $3.9\%$, $3.4\%$ and
$2.3\%$ of their text lies beyond 512 tokens, which is comparable to TREC-COVID. We therefore measure them with
windowing, at the same budget as the exclusion columns, so that a system's no-harm and exclusion results are
computed on the same documents (Table~\ref{tab:strong_noharm}).

All systems but one score lower on ordinary retrieval, and the loss follows neither size nor family. The three dense
retrievers lose $0.032$, $0.033$ and $0.207$, whereas Promptriever gains and falls short on exclusion instead
(Table~\ref{tab:strong}). The four cross-encoders lose between $0.043$ and $0.095$, and the smallest loses the
most. The loss of gte-Qwen2 is more than twice the next largest. In relative terms the loss is largest on
ArguAna for six of the eight systems: its long argumentative queries are those on which an exclusion-tuned
model is most likely to act, which makes this suite the most demanding test of no-harm behaviour for every
system, including ours.

\begin{table*}[t]\centering
\caption{\textbf{Ordinary retrieval per suite, off the shelf $\to$ fine-tuned}, at the same $10{,}000$-word budget as the exclusion columns. Values are nDCG@10 on SciFact, NFCorpus and ArguAna, and their mean. The systems start from different levels, so the change within a system, not its absolute level, measures the cost to ordinary retrieval. The last row compares the frozen index with \excise{}, which re-embeds the shortlist when its detector fires. In the mean column, \textbf{bold} marks the largest decrease and the only increase.}
\label{tab:strong_noharm}
\resizebox{\textwidth}{!}{%
\begin{tabular}{@{}lcccc@{}}
\toprule
\textbf{System} (off\,$\to$\,ft) & \textbf{SciFact} & \textbf{NFCorpus} & \textbf{ArguAna} & \textbf{Mean} \\
\midrule
BGE-large      & $0.746\!\to\!0.698$ & $0.383\!\to\!0.371$ & $0.460\!\to\!0.424$ & $0.530\!\to\!0.498$ \\
E5-large       & $0.715\!\to\!0.634$ & $0.350\!\to\!0.319$ & $0.344\!\to\!0.357$ & $0.470\!\to\!0.436$ \\
gte-Qwen2-1.5B & $0.781\!\to\!0.550$ & $0.393\!\to\!0.248$ & $0.479\!\to\!0.235$ & $\mathbf{0.551\!\to\!0.344}$ \\
Promptriever   & $0.646\!\to\!0.682$ & $0.337\!\to\!0.363$ & $0.379\!\to\!0.418$ & $\mathbf{0.454\!\to\!0.488}$ \\
\midrule
MS-MARCO MiniLM     & $0.689\!\to\!0.551$ & $0.362\!\to\!0.335$ & $0.316\!\to\!0.194$ & $0.456\!\to\!0.360$ \\
bge-reranker-base   & $0.689\!\to\!0.583$ & $0.314\!\to\!0.330$ & $0.180\!\to\!0.143$ & $0.395\!\to\!0.352$ \\
bge-reranker-large  & $0.725\!\to\!0.574$ & $0.344\!\to\!0.355$ & $0.209\!\to\!0.153$ & $0.426\!\to\!0.361$ \\
bge-reranker-v2-m3  & $0.731\!\to\!0.571$ & $0.350\!\to\!0.355$ & $0.281\!\to\!0.196$ & $0.454\!\to\!0.374$ \\
\midrule
Frozen ColBERT $\to$ \excise{} & $0.752\!\to\!0.744$ & $0.381\!\to\!0.381$ & $0.405\!\to\!0.397$ & $0.513\!\to\!0.507$ \\
\bottomrule
\end{tabular}}
\end{table*}

\paragraph{Rehearsal as a mitigation.} These systems were not trained on exclusion alone, which might otherwise explain their loss on ordinary retrieval: close to a quarter of the training triples contain no exclusion. Rehearsal is nevertheless the standard remedy for forgetting, so we tested it on
gte-Qwen2, the system with the largest loss. The model was retrained with an equal number of SciDocs pairs
added and everything else unchanged. SciDocs is an ordinary retrieval corpus that is not among the test or control sets.

Rehearsal helps, but only partly. All three suites improve, from $0.5500$ to $0.5619$ on SciFact, from
$0.2480$ to $0.2794$ on NFCorpus and from $0.2347$ to $0.2544$ on ArguAna, which raises the mean from $0.3442$
to $0.3652$. Relative to the $0.5512$ that the model scored before fine-tuning, rehearsal recovers about a
tenth of the loss.

The remaining cost is shifted rather than removed. The two in-domain collections move in opposite directions:
FiQA rises from $0.5337$ to $0.5625$, while ExcluIR falls from $0.7148$ to $0.6661$. Pooled over both, the
system loses $0.028$ of exclusion success@10 in exchange for the $0.021$ of ordinary retrieval that it
regains. Rehearsal therefore moves a system along the trade-off instead of removing it. \excise{} is not
subject to this trade-off, because it never updates the encoder: its ordinary retrieval, $0.507$, stays close
to the $0.513$ of the frozen backbone. This result has one limitation: it covers a single training setup at a single mixing ratio, so it shows that this mitigation does not avoid the trade-off, not that no mitigation
can.

\paragraph{The off-the-shelf rows.} The discussion above concerns fine-tuned systems. Cross-encoders also have
a cost before any training: re-scoring BGE-large's shortlist changes ArguAna from $0.460$ to $0.180$--$0.316$
(Table~\ref{tab:strong_noharm}), because a reranker without a gate re-scores every query, whether or not it
contains an exclusion. The switch in \excise{} avoids this cost. These rows should also be read in light of
the gate asymmetry described in Appendix~\ref{app:bench}, since the external systems retrieve their own
shortlists.

\section{Complete Results}
\label{app:partE}

\paragraph{Significance of the main results.} For each backbone and collection of Table~\ref{tab:main}, we test
the lead of \excise{} over the strongest baseline with the paired bootstrap of \S\ref{sec:setup}. The lead is
significant at the $0.05$ level in $16$ of the $18$ backbone--collection combinations; the two exceptions are
EDGAR and EUR-Lex on ColBERTv2.

\subsection{Ordinary Retrieval and Boolean Control, in Full}
\label{app:noharm}

Table~\ref{tab:noharm} reports the per-suite no-harm results summarised in \S\ref{sec:results-noharm},
together with the Boolean accuracies. Two points should be kept in mind when reading it. First, the backbones start from different no-harm levels, because each backbone indexes with its own encoder; each row should therefore be compared with the frozen reference of the same backbone, which for Reason is $0.5126$ against
the operator's $0.5072$. Second, on AND and OR the frozen index is the target: a safe operator should leave
these two connectives unchanged, so any change is a cost. On NOT, in contrast, the frozen result is the
failure that this paper addresses.

When the operator runs on a Boolean pair, the soft penalty changes AND and OR without violating
Proposition~\ref{prop:flat}. The proposition guarantees an unchanged ranking only when the spread of the penalty is smaller
than the smallest gap between adjacent candidates. On a genuine conjunction or disjunction the evidence varies
across candidates, so the penalty varies as well and the condition does not hold. These candidate sets also
contain only two or three documents, so they fall under the absolute rule of Appendix~\ref{app:sens}, where no hard demotion takes place and only the soft penalty acts.

\begin{table*}[t]\centering
\caption{\textbf{Ordinary retrieval and Boolean control for every system on every backbone.} Left: nDCG@10 on the three no-harm suites. The aim is to \emph{match} the frozen index, not to exceed it, so no value is bolded. The backbones start from different levels, so each row should be compared with the frozen row of the same backbone. Right: pairwise accuracy on the $834$ scorable BoolQuestions pairs. On AND and OR the aim is again to match the frozen index; only on NOT is a higher value better.}
\label{tab:noharm}
\resizebox{\textwidth}{!}{%
\begin{tabular}{@{}lcccc cccc@{}}
\toprule
 & \multicolumn{4}{c}{\textbf{No harm} (nDCG@10)} & \multicolumn{4}{c}{\textbf{Boolean} (pairwise accuracy)} \\
\cmidrule(lr){2-5}\cmidrule(lr){6-9}
\textbf{System} & SciFact & NFCorpus & ArguAna & Mean & AND & OR & NOT & Mean \\
\midrule
\rowcolor{gray!12}\multicolumn{9}{c}{\textit{Reason-ModernColBERT}} \\
Frozen index \emph{(reference)} & 0.7519 & 0.3812 & 0.4048 & 0.5126 & 0.9459 & 0.9803 & 0.2915 & 0.7392 \\
LoRA fine-tune & 0.7215 & 0.3700 & 0.4339 & 0.5085 & 0.8709 & 0.9331 & 0.7449 & 0.8496 \\
Full fine-tune & 0.6795 & 0.3690 & 0.4266 & 0.4917 & 0.8589 & 0.9173 & 0.6559 & 0.8107 \\
Query-only fine-tune & 0.7271 & 0.3855 & 0.4011 & 0.5046 & 0.8769 & 0.9252 & 0.7409 & 0.8477 \\
\oursrow \textbf{\excise{}} & 0.7438 & 0.3808 & 0.3971 & 0.5072 & 0.8979 & 0.9449 & 0.8907 & 0.9112 \\
\midrule
\rowcolor{gray!12}\multicolumn{9}{c}{\textit{GTE-ModernColBERT-v1}} \\
Frozen index \emph{(reference)} & 0.7621 & 0.3778 & 0.3334 & 0.4911 & 0.9429 & 0.9921 & 0.2915 & 0.7422 \\
LoRA fine-tune & 0.7407 & 0.3704 & 0.4128 & 0.5080 & 0.8649 & 0.9449 & 0.7449 & 0.8516 \\
Full fine-tune & 0.6526 & 0.3542 & 0.3984 & 0.4684 & 0.8649 & 0.9213 & 0.6923 & 0.8262 \\
Query-only fine-tune & 0.6513 & 0.3569 & 0.3229 & 0.4437 & 0.8078 & 0.8780 & 0.7449 & 0.8102 \\
\oursrow \textbf{\excise{}} & 0.7609 & 0.3778 & 0.3341 & 0.4909 & 0.9009 & 0.9606 & 0.9231 & 0.9282 \\
\midrule
\rowcolor{gray!12}\multicolumn{9}{c}{\textit{ColBERTv2.0}} \\
Frozen index \emph{(reference)} & 0.7025 & 0.3464 & 0.3501 & 0.4663 & 0.9309 & 0.9764 & 0.2510 & 0.7194 \\
LoRA fine-tune & 0.6143 & 0.3360 & 0.3357 & 0.4287 & 0.8859 & 0.8740 & 0.7045 & 0.8215 \\
Full fine-tune & 0.5760 & 0.3250 & 0.3215 & 0.4075 & 0.8679 & 0.8819 & 0.6640 & 0.8046 \\
Query-only fine-tune & 0.6191 & 0.3367 & 0.3065 & 0.4208 & 0.8739 & 0.9016 & 0.6802 & 0.8185 \\
\oursrow \textbf{\excise{}} & 0.7029 & 0.3445 & 0.3484 & 0.4653 & 0.8919 & 0.9331 & 0.8421 & 0.8890 \\
\bottomrule
\end{tabular}}
\end{table*}

\subsection{The Component Ablation, in Full}
\label{app:ablation}

Table~\ref{tab:ablation} reports the contribution of each component, as a cumulative block and as a leave-one-out block for each collection. The two blocks are different experiments: a row of the cumulative block omits every later stage, so it should not be read as the removal of a single component.

Three rows deserve comment. First, removing the switch raises exclusion slightly on three of the six
collections and leaves the other three unchanged. We nevertheless keep the switch: every query in these
collections contains an exclusion, so the switch can only withhold the operator from a query that needs it,
whereas its benefit lies in the ordinary queries that it leaves untouched. Second, the soft-penalty configuration gains exclusion at a cost in ranking quality, which the next stage recovers. On the explicit tier of ExcluIR,
xnDCG is $0.596$ for the soft-penalty configuration, against $0.805$ for the soft penalty on frozen vectors, and the next stage raises it to $0.752$; the cue-free and compound tiers show the same pattern (Table~\ref{tab:tm_excluir}). Every cumulative configuration remains above the frozen index, whose xnDCG is $0.368$, $0.243$ and $0.041$ on the three tiers and which
leaves an excluded document in the top ten for $81\%$ of its T1 queries.

Third, the final stage, hard demotion, has little effect on success@10: it costs $0.005$ on ExcluIR and on
FiQA, gains $0.024$ on ESGenius, and adds $0.002$ on average. This is expected, because success@10 counts a
query as solved once no excluded document remains in the top ten, so a stage that moves the last remaining excluded document out of the top ten affects the metric only when it changes that outcome. Its effect is visible in leak, as
\S\ref{sec:method-demotion} describes: hard demotion lowers leak on five of the six collections and raises it
on none, by $0.022$ on ESGenius and $0.006$ on average, while mean xnDCG changes by $+0.0003$.

\begin{table*}[t]\centering
\caption{\textbf{Component ablation on Reason-ModernColBERT}: exclusion success@10 per collection on the test set. The upper block adds the stages one at a time, in the order in which the operator applies them. The lower block removes one component from the deployed system. The rows \excise{}$_{\text{re-emb}}$ and \emph{$-$ demotion} are the same configuration. \emph{Soft penalty on frozen vectors} applies only the soft penalty of \excise{}, with no re-embedding, no floor or cutoff, and no hard demotion. The highlighted row is the deployed operator.}
\label{tab:ablation}
\resizebox{\textwidth}{!}{%
\begin{tabular}{@{}llcccccc@{}}
\toprule
\textbf{Configuration} & \textbf{what it adds / removes} & ExcluIR & FiQA & TREC-C & ESGen & EDGAR & EUR-Lex \\
\midrule
\rowcolor{gray!12}\multicolumn{8}{c}{\textit{cumulative, in the order the operator applies its stages}} \\
Frozen ColBERT & \emph{reference} & 0.1354 & 0.1827 & 0.1866 & 0.1699 & 0.3031 & 0.2806 \\
\excise{}$_{\text{re-emb}}$ & re-embed the shortlist & 0.7058 & 0.5625 & 0.6589 & 0.3350 & 0.5609 & 0.6429 \\
\excise{}$_{\text{soft}}$ & $+$ soft penalty & 0.7924 & 0.7404 & 0.8192 & 0.5680 & 0.6374 & 0.7092 \\
\excise{}$_{\text{cut}}$ & $+$ evidence floor, relative cutoff & 0.8556 & 0.7596 & 0.8455 & 0.6311 & 0.6799 & 0.7449 \\
\oursrow \textbf{\excise{}} \textit{(deployed)} & $+$ hard demotion, guard & 0.8502 & 0.7548 & 0.8455 & 0.6553 & 0.6771 & 0.7449 \\
\midrule
\rowcolor{gray!12}\multicolumn{8}{c}{\textit{leave one out of the deployed system}} \\
\;$-$ soft penalty & re-embed and hard demotion & 0.8538 & 0.7115 & 0.7959 & 0.5922 & 0.6459 & 0.6786 \\
\;$-$ switch & re-embeds every query & 0.8520 & 0.7692 & 0.8513 & 0.6553 & 0.6771 & 0.7449 \\
\;$-$ re-embedding & all demotion stages, frozen vectors & 0.8303 & 0.6587 & 0.7259 & 0.5340 & 0.6147 & 0.6480 \\
\;$-$ demotion & re-embed only & 0.7058 & 0.5625 & 0.6589 & 0.3350 & 0.5609 & 0.6429 \\
\midrule
\rowcolor{gray!12}\multicolumn{8}{c}{\textit{the demotion rule alone}} \\
Soft penalty on frozen vectors & no re-embedding, no floor or cutoff, no hard demotion & 0.8123 & 0.7788 & 0.7901 & 0.6068 & 0.6714 & 0.7449 \\
\bottomrule
\end{tabular}}
\end{table*}

\subsection{Per-Tier Results}
\label{app:pertier}

The preceding tables report one collection at a time. Two further sets of tables break the results down by
difficulty tier. Table~\ref{tab:pertier} reports exclusion success@10 and leak for the deployed operator on
every backbone and collection. Tables~\ref{tab:tm_excluir}--\ref{tab:tm_eurlex} report all five metrics per
tier for every system and ablation configuration on Reason. Among them, xnDCG is nDCG@10 with a gain of $-\beta$ for each excluded document in the top ten, so it can be negative. Across Tables~\ref{tab:tm_excluir}--\ref{tab:tm_eurlex}, the deployed operator leads all
four baselines on both success@10 and xnDCG in almost all of the $192$ comparisons they contain (six collections, four baselines, two metrics, and the full collection plus each of its three tiers). The per-tier breakdown
matters because a lead that holds for a collection as a whole could fail on one of its tiers.

The two metrics do not always agree for other configurations. Success@10 is binary: the gold must appear in
the top ten and no excluded document may appear there, but the rank of the gold does not matter, and one
leaked document counts as much as nine. xnDCG grades both aspects, rewarding a higher rank for the gold and
penalising each excluded document that remains, so a configuration can lead on one metric and trail on the
other. On the explicit tier of ExcluIR, \excise{}$_{\text{soft}}$ solves more queries than the LoRA fine-tune
($0.764$ against $0.714$) with a lower leak ($0.005$ against $0.206$), yet ranks the gold lower (nDCG $0.598$
against $0.818$) and trails on xnDCG ($0.596$ against $0.725$). A single metric would have hidden part of this
behaviour, which is why both metrics are reported.

The compound tier, not the cue-free tier, is the most difficult. T3 has the lowest success in $14$ of the $18$
backbone--collection combinations and the highest leak in $9$, against $2$ and $6$ for T2. A compound query
excludes two topics and has a separate confusable negative for each (Appendix~\ref{app:bench}), so the operator
must demote two documents for a single query. The residual error is therefore associated with compound
exclusions rather than with the absence of a lexical cue.

\begin{table}[t]\centering\small
\caption{Exclusion success@10 and leak (lower is better) of the deployed operator, by difficulty tier. T1 states the exclusion with a cue, T2 contains no cue, and T3 excludes two topics. \textbf{Bold} marks the weakest tier in each row by success@10, together with its leak.}
\label{tab:pertier}
\resizebox{\columnwidth}{!}{%
\begin{tabular}{@{}l cccc c cccc@{}}
\toprule
 & \multicolumn{4}{c}{\textbf{success@10}} & & \multicolumn{4}{c}{\textbf{leak} $\downarrow$} \\
\cmidrule(lr){2-5}\cmidrule(lr){7-10}
\textbf{Collection} & all & T1 & T2 & T3 & & all & T1 & T2 & T3 \\
\midrule
\rowcolor{gray!12}\multicolumn{10}{c}{\textit{Reason-ModernColBERT}} \\
ExcluIR & 0.850 & 0.869 & \textbf{0.837} & 0.842 & & 0.021 & 0.005 & \textbf{0.023} & 0.036 \\
FiQA & 0.755 & \textbf{0.727} & 0.806 & 0.739 & & 0.120 & \textbf{0.156} & 0.129 & 0.072 \\
TREC-COVID & 0.846 & 0.883 & 0.845 & \textbf{0.796} & & 0.073 & 0.044 & 0.087 & \textbf{0.097} \\
ESGenius & 0.655 & 0.720 & 0.645 & \textbf{0.581} & & 0.153 & 0.146 & 0.129 & \textbf{0.185} \\
EDGAR & 0.677 & 0.790 & 0.640 & \textbf{0.602} & & 0.085 & 0.034 & 0.135 & \textbf{0.089} \\
EUR-Lex & 0.745 & 0.797 & 0.763 & \textbf{0.655} & & 0.084 & 0.076 & 0.085 & \textbf{0.095} \\
\rowcolor{gray!12}\multicolumn{10}{c}{\textit{GTE-ModernColBERT-v1}} \\
ExcluIR & 0.836 & \textbf{0.824} & 0.826 & 0.858 & & 0.015 & \textbf{0.010} & 0.023 & 0.014 \\
FiQA & 0.716 & 0.701 & 0.790 & \textbf{0.667} & & 0.118 & 0.156 & 0.113 & \textbf{0.080} \\
TREC-COVID & 0.808 & 0.876 & 0.777 & \textbf{0.748} & & 0.071 & 0.051 & 0.107 & \textbf{0.063} \\
ESGenius & 0.670 & 0.646 & 0.726 & \textbf{0.645} & & 0.136 & 0.122 & 0.145 & \textbf{0.145} \\
EDGAR & 0.603 & 0.698 & 0.595 & \textbf{0.520} & & 0.078 & 0.050 & 0.108 & \textbf{0.077} \\
EUR-Lex & 0.699 & 0.722 & 0.780 & \textbf{0.586} & & 0.105 & 0.101 & 0.068 & \textbf{0.147} \\
\rowcolor{gray!12}\multicolumn{10}{c}{\textit{ColBERTv2.0}} \\
ExcluIR & 0.843 & 0.864 & \textbf{0.826} & 0.836 & & 0.028 & 0.025 & \textbf{0.023} & 0.036 \\
FiQA & 0.707 & 0.675 & 0.790 & \textbf{0.667} & & 0.106 & 0.169 & 0.065 & \textbf{0.072} \\
TREC-COVID & 0.735 & 0.810 & 0.738 & \textbf{0.631} & & 0.093 & 0.080 & 0.126 & \textbf{0.078} \\
ESGenius & 0.626 & 0.707 & 0.645 & \textbf{0.500} & & 0.189 & 0.110 & 0.226 & \textbf{0.258} \\
EDGAR & 0.524 & 0.622 & 0.550 & \textbf{0.407} & & 0.127 & 0.118 & 0.126 & \textbf{0.138} \\
EUR-Lex & 0.617 & 0.658 & 0.644 & \textbf{0.535} & & 0.115 & 0.114 & 0.085 & \textbf{0.147} \\
\bottomrule
\end{tabular}}
\end{table}

\begin{table}[t]\centering\small
\caption{\textbf{ExcluIR: every metric by tier}, Reason-ModernColBERT. \textbf{succ}@10, \textbf{hit}@10 and \textbf{leak} are defined in \S\ref{sec:nogo}, and \textbf{xnDCG}$_{\beta=1}$ in \S\ref{sec:setup}. \textbf{all} is the collection as a whole; the three tiers partition it. \excise{}$_{\text{demote}}$ is the soft penalty on frozen vectors (Table~\ref{tab:ablation}).}
\label{tab:tm_excluir}
\resizebox{\columnwidth}{!}{%
\begin{tabular}{@{}llccccc@{}}
\toprule
\textbf{System} & & \textbf{succ} & \textbf{nDCG} & \textbf{hit} & \textbf{leak}$\downarrow$ & \textbf{xnDCG} \\
\midrule
Frozen & \textbf{all} & 0.135 & 0.836 & 0.939 & 0.788 & 0.221 \\
 & T1 & 0.176 & 0.893 & 0.975 & 0.814 & 0.368 \\
 & T2 & 0.076 & 0.840 & 0.930 & 0.913 & 0.243 \\
 & T3 & 0.147 & 0.770 & 0.907 & 0.642 & 0.041 \\
\addlinespace[1.5pt]
LoRA-FT$^{\dagger}$ & \textbf{all} & 0.682 & 0.801 & 0.890 & 0.191 & 0.694 \\
 & T1 & 0.714 & 0.818 & 0.910 & 0.206 & 0.725 \\
 & T2 & 0.657 & 0.796 & 0.884 & 0.250 & 0.662 \\
 & T3 & 0.672 & 0.787 & 0.874 & 0.120 & 0.690 \\
\addlinespace[1.5pt]
Full-FT$^{\dagger}$ & \textbf{all} & 0.581 & 0.608 & 0.718 & 0.144 & 0.524 \\
 & T1 & 0.578 & 0.625 & 0.739 & 0.196 & 0.523 \\
 & T2 & 0.587 & 0.627 & 0.721 & 0.157 & 0.543 \\
 & T3 & 0.579 & 0.571 & 0.694 & 0.077 & 0.507 \\
\addlinespace[1.5pt]
Query-only$^{\ddagger}$ & \textbf{all} & 0.666 & 0.875 & 0.939 & 0.247 & 0.740 \\
 & T1 & 0.663 & 0.898 & 0.965 & 0.307 & 0.757 \\
 & T2 & 0.680 & 0.861 & 0.924 & 0.273 & 0.728 \\
 & T3 & 0.656 & 0.865 & 0.923 & 0.158 & 0.732 \\
\addlinespace[1.5pt]
\excise{}$_{\text{demote}}$ & \textbf{all} & 0.812 & 0.772 & 0.852 & 0.035 & 0.742 \\
 & T1 & 0.874 & 0.808 & 0.884 & 0.010 & 0.805 \\
 & T2 & 0.808 & 0.769 & 0.849 & 0.047 & 0.743 \\
 & T3 & 0.749 & 0.735 & 0.820 & 0.052 & 0.671 \\
\addlinespace[1.5pt]
\excise{}$_{\text{soft}}$ & \textbf{all} & 0.792 & 0.654 & 0.809 & 0.014 & 0.643 \\
 & T1 & 0.764 & 0.598 & 0.764 & 0.005 & 0.596 \\
 & T2 & 0.802 & 0.694 & 0.826 & 0.023 & 0.677 \\
 & T3 & 0.814 & 0.678 & 0.842 & 0.014 & 0.662 \\
\addlinespace[1.5pt]
\excise{}$_{\text{cut}}$ & \textbf{all} & 0.856 & 0.786 & 0.884 & 0.023 & 0.769 \\
 & T1 & 0.879 & 0.754 & 0.879 & 0.005 & 0.752 \\
 & T2 & 0.843 & 0.789 & 0.866 & 0.029 & 0.770 \\
 & T3 & 0.842 & 0.818 & 0.907 & 0.038 & 0.786 \\
\addlinespace[1.5pt]
\oursrow \textbf{\excise{}} & \textbf{all} & 0.850 & 0.784 & 0.877 & 0.021 & 0.768 \\
\oursrow  & T1 & 0.869 & 0.751 & 0.869 & 0.005 & 0.750 \\
\oursrow  & T2 & 0.837 & 0.787 & 0.860 & 0.023 & 0.770 \\
\oursrow  & T3 & 0.842 & 0.816 & 0.902 & 0.036 & 0.786 \\
\addlinespace[1.5pt]
\bottomrule
\end{tabular}}
\end{table}

\begin{table}[t]\centering\small
\caption{\textbf{FiQA: every metric by tier}, Reason-ModernColBERT. \textbf{succ}@10, \textbf{hit}@10 and \textbf{leak} are defined in \S\ref{sec:nogo}, and \textbf{xnDCG}$_{\beta=1}$ in \S\ref{sec:setup}. \textbf{all} is the collection as a whole; the three tiers partition it. \excise{}$_{\text{demote}}$ is the soft penalty on frozen vectors (Table~\ref{tab:ablation}).}
\label{tab:tm_fiqa}
\resizebox{\columnwidth}{!}{%
\begin{tabular}{@{}llccccc@{}}
\toprule
\textbf{System} & & \textbf{succ} & \textbf{nDCG} & \textbf{hit} & \textbf{leak}$\downarrow$ & \textbf{xnDCG} \\
\midrule
Frozen & \textbf{all} & 0.183 & 0.743 & 0.889 & 0.731 & 0.170 \\
 & T1 & 0.117 & 0.756 & 0.909 & 0.857 & 0.153 \\
 & T2 & 0.274 & 0.801 & 0.919 & 0.726 & 0.352 \\
 & T3 & 0.174 & 0.676 & 0.841 & 0.594 & 0.024 \\
\addlinespace[1.5pt]
LoRA-FT$^{\dagger}$ & \textbf{all} & 0.514 & 0.717 & 0.870 & 0.358 & 0.467 \\
 & T1 & 0.494 & 0.724 & 0.896 & 0.442 & 0.456 \\
 & T2 & 0.565 & 0.731 & 0.887 & 0.387 & 0.510 \\
 & T3 & 0.493 & 0.697 & 0.826 & 0.239 & 0.439 \\
\addlinespace[1.5pt]
Full-FT$^{\dagger}$ & \textbf{all} & 0.543 & 0.659 & 0.827 & 0.286 & 0.460 \\
 & T1 & 0.546 & 0.725 & 0.883 & 0.364 & 0.493 \\
 & T2 & 0.613 & 0.649 & 0.839 & 0.274 & 0.511 \\
 & T3 & 0.478 & 0.594 & 0.754 & 0.210 & 0.378 \\
\addlinespace[1.5pt]
Query-only$^{\ddagger}$ & \textbf{all} & 0.524 & 0.735 & 0.885 & 0.382 & 0.496 \\
 & T1 & 0.494 & 0.753 & 0.909 & 0.481 & 0.502 \\
 & T2 & 0.597 & 0.782 & 0.919 & 0.387 & 0.583 \\
 & T3 & 0.493 & 0.671 & 0.826 & 0.268 & 0.410 \\
\addlinespace[1.5pt]
\excise{}$_{\text{demote}}$ & \textbf{all} & 0.779 & 0.761 & 0.875 & 0.094 & 0.697 \\
 & T1 & 0.779 & 0.765 & 0.857 & 0.091 & 0.708 \\
 & T2 & 0.855 & 0.805 & 0.935 & 0.097 & 0.758 \\
 & T3 & 0.710 & 0.716 & 0.841 & 0.094 & 0.631 \\
\addlinespace[1.5pt]
\excise{}$_{\text{soft}}$ & \textbf{all} & 0.740 & 0.613 & 0.803 & 0.077 & 0.572 \\
 & T1 & 0.701 & 0.583 & 0.766 & 0.091 & 0.525 \\
 & T2 & 0.758 & 0.609 & 0.855 & 0.113 & 0.570 \\
 & T3 & 0.768 & 0.649 & 0.797 & 0.029 & 0.627 \\
\addlinespace[1.5pt]
\excise{}$_{\text{cut}}$ & \textbf{all} & 0.760 & 0.728 & 0.885 & 0.127 & 0.650 \\
 & T1 & 0.740 & 0.720 & 0.896 & 0.169 & 0.631 \\
 & T2 & 0.806 & 0.774 & 0.919 & 0.129 & 0.701 \\
 & T3 & 0.739 & 0.694 & 0.841 & 0.080 & 0.624 \\
\addlinespace[1.5pt]
\oursrow \textbf{\excise{}} & \textbf{all} & 0.755 & 0.722 & 0.870 & 0.120 & 0.648 \\
\oursrow  & T1 & 0.727 & 0.705 & 0.857 & 0.156 & 0.621 \\
\oursrow  & T2 & 0.806 & 0.774 & 0.919 & 0.129 & 0.701 \\
\oursrow  & T3 & 0.739 & 0.694 & 0.841 & 0.072 & 0.630 \\
\addlinespace[1.5pt]
\bottomrule
\end{tabular}}
\end{table}

\begin{table}[t]\centering\small
\caption{\textbf{TREC-COVID: every metric by tier}, Reason-ModernColBERT. \textbf{succ}@10, \textbf{hit}@10 and \textbf{leak} are defined in \S\ref{sec:nogo}, and \textbf{xnDCG}$_{\beta=1}$ in \S\ref{sec:setup}. \textbf{all} is the collection as a whole; the three tiers partition it. \excise{}$_{\text{demote}}$ is the soft penalty on frozen vectors (Table~\ref{tab:ablation}).}
\label{tab:tm_trec-covid}
\resizebox{\columnwidth}{!}{%
\begin{tabular}{@{}llccccc@{}}
\toprule
\textbf{System} & & \textbf{succ} & \textbf{nDCG} & \textbf{hit} & \textbf{leak}$\downarrow$ & \textbf{xnDCG} \\
\midrule
Frozen & \textbf{all} & 0.187 & 0.747 & 0.895 & 0.736 & 0.215 \\
 & T1 & 0.182 & 0.770 & 0.934 & 0.818 & 0.258 \\
 & T2 & 0.184 & 0.790 & 0.922 & 0.796 & 0.333 \\
 & T3 & 0.194 & 0.673 & 0.816 & 0.568 & 0.041 \\
\addlinespace[1.5pt]
LoRA-FT$^{\dagger}$ & \textbf{all} & 0.644 & 0.811 & 0.933 & 0.273 & 0.657 \\
 & T1 & 0.693 & 0.828 & 0.920 & 0.248 & 0.714 \\
 & T2 & 0.573 & 0.814 & 0.951 & 0.379 & 0.642 \\
 & T3 & 0.651 & 0.787 & 0.932 & 0.199 & 0.598 \\
\addlinespace[1.5pt]
Full-FT$^{\dagger}$ & \textbf{all} & 0.618 & 0.712 & 0.851 & 0.227 & 0.584 \\
 & T1 & 0.628 & 0.739 & 0.861 & 0.270 & 0.611 \\
 & T2 & 0.641 & 0.749 & 0.893 & 0.252 & 0.623 \\
 & T3 & 0.583 & 0.639 & 0.796 & 0.146 & 0.510 \\
\addlinespace[1.5pt]
Query-only$^{\ddagger}$ & \textbf{all} & 0.682 & 0.796 & 0.921 & 0.232 & 0.666 \\
 & T1 & 0.730 & 0.829 & 0.949 & 0.234 & 0.717 \\
 & T2 & 0.651 & 0.812 & 0.932 & 0.291 & 0.685 \\
 & T3 & 0.651 & 0.738 & 0.874 & 0.170 & 0.580 \\
\addlinespace[1.5pt]
\excise{}$_{\text{demote}}$ & \textbf{all} & 0.790 & 0.739 & 0.860 & 0.073 & 0.680 \\
 & T1 & 0.905 & 0.807 & 0.942 & 0.044 & 0.782 \\
 & T2 & 0.777 & 0.729 & 0.845 & 0.087 & 0.674 \\
 & T3 & 0.650 & 0.658 & 0.767 & 0.097 & 0.551 \\
\addlinespace[1.5pt]
\excise{}$_{\text{soft}}$ & \textbf{all} & 0.819 & 0.720 & 0.857 & 0.047 & 0.688 \\
 & T1 & 0.898 & 0.790 & 0.912 & 0.022 & 0.777 \\
 & T2 & 0.767 & 0.687 & 0.835 & 0.078 & 0.646 \\
 & T3 & 0.767 & 0.658 & 0.806 & 0.049 & 0.613 \\
\addlinespace[1.5pt]
\excise{}$_{\text{cut}}$ & \textbf{all} & 0.846 & 0.802 & 0.927 & 0.074 & 0.750 \\
 & T1 & 0.883 & 0.829 & 0.920 & 0.044 & 0.807 \\
 & T2 & 0.854 & 0.791 & 0.932 & 0.087 & 0.740 \\
 & T3 & 0.786 & 0.778 & 0.932 & 0.102 & 0.684 \\
\addlinespace[1.5pt]
\oursrow \textbf{\excise{}} & \textbf{all} & 0.846 & 0.801 & 0.924 & 0.073 & 0.750 \\
\oursrow  & T1 & 0.883 & 0.829 & 0.920 & 0.044 & 0.807 \\
\oursrow  & T2 & 0.845 & 0.788 & 0.922 & 0.087 & 0.736 \\
\oursrow  & T3 & 0.796 & 0.778 & 0.932 & 0.097 & 0.687 \\
\addlinespace[1.5pt]
\bottomrule
\end{tabular}}
\end{table}

\begin{table}[t]\centering\small
\caption{\textbf{ESGenius: every metric by tier}, Reason-ModernColBERT. \textbf{succ}@10, \textbf{hit}@10 and \textbf{leak} are defined in \S\ref{sec:nogo}, and \textbf{xnDCG}$_{\beta=1}$ in \S\ref{sec:setup}. \textbf{all} is the collection as a whole; the three tiers partition it. \excise{}$_{\text{demote}}$ is the soft penalty on frozen vectors (Table~\ref{tab:ablation}).}
\label{tab:tm_esgenius}
\resizebox{\columnwidth}{!}{%
\begin{tabular}{@{}llccccc@{}}
\toprule
\textbf{System} & & \textbf{succ} & \textbf{nDCG} & \textbf{hit} & \textbf{leak}$\downarrow$ & \textbf{xnDCG} \\
\midrule
Frozen & \textbf{all} & 0.170 & 0.579 & 0.801 & 0.723 & -0.019 \\
 & T1 & 0.146 & 0.605 & 0.829 & 0.829 & -0.001 \\
 & T2 & 0.226 & 0.577 & 0.758 & 0.661 & 0.162 \\
 & T3 & 0.145 & 0.546 & 0.806 & 0.645 & -0.225 \\
\addlinespace[1.5pt]
LoRA-FT$^{\dagger}$ & \textbf{all} & 0.335 & 0.711 & 0.864 & 0.551 & 0.343 \\
 & T1 & 0.317 & 0.690 & 0.841 & 0.622 & 0.313 \\
 & T2 & 0.371 & 0.754 & 0.871 & 0.565 & 0.468 \\
 & T3 & 0.323 & 0.696 & 0.887 & 0.444 & 0.258 \\
\addlinespace[1.5pt]
Full-FT$^{\dagger}$ & \textbf{all} & 0.354 & 0.595 & 0.801 & 0.468 & 0.281 \\
 & T1 & 0.305 & 0.623 & 0.793 & 0.585 & 0.285 \\
 & T2 & 0.419 & 0.577 & 0.806 & 0.435 & 0.352 \\
 & T3 & 0.355 & 0.576 & 0.806 & 0.347 & 0.204 \\
\addlinespace[1.5pt]
Query-only$^{\ddagger}$ & \textbf{all} & 0.364 & 0.678 & 0.845 & 0.505 & 0.332 \\
 & T1 & 0.366 & 0.679 & 0.854 & 0.573 & 0.354 \\
 & T2 & 0.419 & 0.684 & 0.823 & 0.468 & 0.429 \\
 & T3 & 0.307 & 0.670 & 0.855 & 0.452 & 0.206 \\
\addlinespace[1.5pt]
\excise{}$_{\text{demote}}$ & \textbf{all} & 0.607 & 0.594 & 0.782 & 0.199 & 0.450 \\
 & T1 & 0.646 & 0.690 & 0.841 & 0.244 & 0.568 \\
 & T2 & 0.581 & 0.478 & 0.677 & 0.145 & 0.401 \\
 & T3 & 0.581 & 0.584 & 0.806 & 0.194 & 0.344 \\
\addlinespace[1.5pt]
\excise{}$_{\text{soft}}$ & \textbf{all} & 0.568 & 0.531 & 0.694 & 0.121 & 0.433 \\
 & T1 & 0.658 & 0.591 & 0.756 & 0.122 & 0.521 \\
 & T2 & 0.532 & 0.453 & 0.613 & 0.097 & 0.390 \\
 & T3 & 0.484 & 0.529 & 0.694 & 0.145 & 0.361 \\
\addlinespace[1.5pt]
\excise{}$_{\text{cut}}$ & \textbf{all} & 0.631 & 0.668 & 0.835 & 0.175 & 0.531 \\
 & T1 & 0.683 & 0.700 & 0.854 & 0.183 & 0.584 \\
 & T2 & 0.629 & 0.628 & 0.758 & 0.145 & 0.552 \\
 & T3 & 0.565 & 0.667 & 0.887 & 0.194 & 0.442 \\
\addlinespace[1.5pt]
\oursrow \textbf{\excise{}} & \textbf{all} & 0.655 & 0.666 & 0.825 & 0.153 & 0.539 \\
\oursrow  & T1 & 0.720 & 0.694 & 0.841 & 0.146 & 0.594 \\
\oursrow  & T2 & 0.645 & 0.628 & 0.758 & 0.129 & 0.557 \\
\oursrow  & T3 & 0.581 & 0.667 & 0.871 & 0.185 & 0.449 \\
\addlinespace[1.5pt]
\bottomrule
\end{tabular}}
\end{table}

\begin{table}[t]\centering\small
\caption{\textbf{EDGAR: every metric by tier}, Reason-ModernColBERT. \textbf{succ}@10, \textbf{hit}@10 and \textbf{leak} are defined in \S\ref{sec:nogo}, and \textbf{xnDCG}$_{\beta=1}$ in \S\ref{sec:setup}. \textbf{all} is the collection as a whole; the three tiers partition it. \excise{}$_{\text{demote}}$ is the soft penalty on frozen vectors (Table~\ref{tab:ablation}).}
\label{tab:tm_edgar}
\resizebox{\columnwidth}{!}{%
\begin{tabular}{@{}llccccc@{}}
\toprule
\textbf{System} & & \textbf{succ} & \textbf{nDCG} & \textbf{hit} & \textbf{leak}$\downarrow$ & \textbf{xnDCG} \\
\midrule
Frozen & \textbf{all} & 0.303 & 0.503 & 0.739 & 0.531 & 0.038 \\
 & T1 & 0.311 & 0.525 & 0.782 & 0.580 & 0.129 \\
 & T2 & 0.342 & 0.572 & 0.811 & 0.595 & 0.126 \\
 & T3 & 0.260 & 0.420 & 0.634 & 0.427 & -0.129 \\
\addlinespace[1.5pt]
LoRA-FT$^{\dagger}$ & \textbf{all} & 0.530 & 0.520 & 0.697 & 0.187 & 0.387 \\
 & T1 & 0.597 & 0.572 & 0.748 & 0.168 & 0.493 \\
 & T2 & 0.505 & 0.490 & 0.676 & 0.261 & 0.317 \\
 & T3 & 0.488 & 0.496 & 0.667 & 0.138 & 0.349 \\
\addlinespace[1.5pt]
Full-FT$^{\dagger}$ & \textbf{all} & 0.501 & 0.475 & 0.640 & 0.143 & 0.363 \\
 & T1 & 0.504 & 0.498 & 0.639 & 0.143 & 0.424 \\
 & T2 & 0.568 & 0.492 & 0.703 & 0.180 & 0.364 \\
 & T3 & 0.439 & 0.436 & 0.585 & 0.110 & 0.303 \\
\addlinespace[1.5pt]
Query-only$^{\ddagger}$ & \textbf{all} & 0.501 & 0.534 & 0.731 & 0.237 & 0.388 \\
 & T1 & 0.546 & 0.545 & 0.723 & 0.202 & 0.462 \\
 & T2 & 0.514 & 0.538 & 0.748 & 0.306 & 0.361 \\
 & T3 & 0.447 & 0.521 & 0.724 & 0.207 & 0.340 \\
\addlinespace[1.5pt]
\excise{}$_{\text{demote}}$ & \textbf{all} & 0.671 & 0.533 & 0.725 & 0.057 & 0.480 \\
 & T1 & 0.781 & 0.591 & 0.798 & 0.017 & 0.580 \\
 & T2 & 0.658 & 0.564 & 0.712 & 0.081 & 0.511 \\
 & T3 & 0.577 & 0.450 & 0.667 & 0.073 & 0.356 \\
\addlinespace[1.5pt]
\excise{}$_{\text{soft}}$ & \textbf{all} & 0.637 & 0.490 & 0.686 & 0.044 & 0.460 \\
 & T1 & 0.698 & 0.546 & 0.723 & 0.025 & 0.532 \\
 & T2 & 0.640 & 0.466 & 0.667 & 0.054 & 0.432 \\
 & T3 & 0.577 & 0.458 & 0.667 & 0.053 & 0.414 \\
\addlinespace[1.5pt]
\excise{}$_{\text{cut}}$ & \textbf{all} & 0.680 & 0.578 & 0.773 & 0.086 & 0.519 \\
 & T1 & 0.798 & 0.629 & 0.832 & 0.034 & 0.611 \\
 & T2 & 0.640 & 0.554 & 0.739 & 0.135 & 0.478 \\
 & T3 & 0.602 & 0.549 & 0.748 & 0.093 & 0.468 \\
\addlinespace[1.5pt]
\oursrow \textbf{\excise{}} & \textbf{all} & 0.677 & 0.573 & 0.768 & 0.085 & 0.516 \\
\oursrow  & T1 & 0.790 & 0.623 & 0.824 & 0.034 & 0.605 \\
\oursrow  & T2 & 0.640 & 0.552 & 0.739 & 0.135 & 0.475 \\
\oursrow  & T3 & 0.602 & 0.544 & 0.740 & 0.089 & 0.465 \\
\addlinespace[1.5pt]
\bottomrule
\end{tabular}}
\end{table}

\begin{table}[t]\centering\small
\caption{\textbf{EUR-Lex: every metric by tier}, Reason-ModernColBERT. \textbf{succ}@10, \textbf{hit}@10 and \textbf{leak} are defined in \S\ref{sec:nogo}, and \textbf{xnDCG}$_{\beta=1}$ in \S\ref{sec:setup}. \textbf{all} is the collection as a whole; the three tiers partition it. \excise{}$_{\text{demote}}$ is the soft penalty on frozen vectors (Table~\ref{tab:ablation}).}
\label{tab:tm_eurlex}
\resizebox{\columnwidth}{!}{%
\begin{tabular}{@{}llccccc@{}}
\toprule
\textbf{System} & & \textbf{succ} & \textbf{nDCG} & \textbf{hit} & \textbf{leak}$\downarrow$ & \textbf{xnDCG} \\
\midrule
Frozen & \textbf{all} & 0.281 & 0.516 & 0.760 & 0.587 & 0.039 \\
 & T1 & 0.317 & 0.554 & 0.835 & 0.633 & 0.110 \\
 & T2 & 0.322 & 0.525 & 0.763 & 0.559 & 0.143 \\
 & T3 & 0.190 & 0.457 & 0.655 & 0.552 & -0.164 \\
\addlinespace[1.5pt]
LoRA-FT$^{\dagger}$ & \textbf{all} & 0.571 & 0.580 & 0.750 & 0.222 & 0.450 \\
 & T1 & 0.633 & 0.594 & 0.785 & 0.241 & 0.463 \\
 & T2 & 0.593 & 0.562 & 0.712 & 0.186 & 0.484 \\
 & T3 & 0.466 & 0.581 & 0.741 & 0.233 & 0.397 \\
\addlinespace[1.5pt]
Full-FT$^{\dagger}$ & \textbf{all} & 0.475 & 0.466 & 0.612 & 0.181 & 0.345 \\
 & T1 & 0.532 & 0.494 & 0.671 & 0.241 & 0.352 \\
 & T2 & 0.492 & 0.437 & 0.576 & 0.119 & 0.372 \\
 & T3 & 0.379 & 0.457 & 0.569 & 0.164 & 0.307 \\
\addlinespace[1.5pt]
Query-only$^{\ddagger}$ & \textbf{all} & 0.592 & 0.586 & 0.811 & 0.219 & 0.457 \\
 & T1 & 0.608 & 0.562 & 0.823 & 0.241 & 0.439 \\
 & T2 & 0.627 & 0.546 & 0.746 & 0.169 & 0.482 \\
 & T3 & 0.535 & 0.661 & 0.862 & 0.241 & 0.457 \\
\addlinespace[1.5pt]
\excise{}$_{\text{demote}}$ & \textbf{all} & 0.745 & 0.583 & 0.811 & 0.071 & 0.532 \\
 & T1 & 0.759 & 0.565 & 0.835 & 0.101 & 0.499 \\
 & T2 & 0.763 & 0.588 & 0.780 & 0.034 & 0.566 \\
 & T3 & 0.707 & 0.601 & 0.810 & 0.069 & 0.542 \\
\addlinespace[1.5pt]
\excise{}$_{\text{soft}}$ & \textbf{all} & 0.709 & 0.515 & 0.735 & 0.033 & 0.488 \\
 & T1 & 0.759 & 0.528 & 0.797 & 0.051 & 0.490 \\
 & T2 & 0.695 & 0.484 & 0.695 & 0.017 & 0.467 \\
 & T3 & 0.655 & 0.530 & 0.690 & 0.026 & 0.508 \\
\addlinespace[1.5pt]
\excise{}$_{\text{cut}}$ & \textbf{all} & 0.745 & 0.612 & 0.827 & 0.084 & 0.546 \\
 & T1 & 0.797 & 0.617 & 0.861 & 0.076 & 0.558 \\
 & T2 & 0.763 & 0.602 & 0.814 & 0.085 & 0.554 \\
 & T3 & 0.655 & 0.617 & 0.793 & 0.095 & 0.523 \\
\addlinespace[1.5pt]
\oursrow \textbf{\excise{}} & \textbf{all} & 0.745 & 0.613 & 0.827 & 0.084 & 0.547 \\
\oursrow  & T1 & 0.797 & 0.617 & 0.861 & 0.076 & 0.558 \\
\oursrow  & T2 & 0.763 & 0.603 & 0.814 & 0.085 & 0.555 \\
\oursrow  & T3 & 0.655 & 0.617 & 0.793 & 0.095 & 0.523 \\
\addlinespace[1.5pt]
\bottomrule
\end{tabular}}
\end{table}

\section{Extended Related Work}
\label{app:partF}
\label{app:related}

Table~\ref{tab:comparison} compares the main families of approaches along the dimensions that matter for
deployment. The paragraphs below discuss each family and state its costs as well as its capabilities.

\begin{table}[!htb]\centering
\caption{\textbf{How approaches to exclusion compare} (\cmark\ full, $\sim$ partial, \xmark\ absent), summarising the measured tables. Each row is an approach as it is ordinarily deployed. In the no-harm column, \cmark\ means that ordinary retrieval is unchanged, $\sim$ that it changes slightly, and \xmark\ that it degrades. \textbf{No heavy query model} means no joint forward pass per candidate and no multi-billion-parameter encoder at query time.}
\label{tab:comparison}
\setlength{\tabcolsep}{5pt}
\resizebox{\columnwidth}{!}{%
\begin{tabular}{@{}lccccc@{}}
\toprule
& \multicolumn{2}{c}{\textbf{Exclusion}} & \multicolumn{3}{c}{\textbf{Deployment}} \\
\cmidrule(lr){2-3}\cmidrule(lr){4-6}
\textbf{Approach} & Explicit & Cue-free & \makecell{No\\re-index} & \makecell{No heavy\\query model} & \makecell{No\\harm} \\
\midrule
Dense retriever~{\scriptsize\citep{karpukhin2020dpr,wang2022e5}}       & \xmark   & \xmark   & \cmark & \cmark & \cmark \\
Cross-encoder reranker~{\scriptsize\citep{nogueira2019passage}}         & \xmark   & \xmark   & \cmark & \xmark & $\sim$ \\
\quad\emph{fine-tuned to exclude}                                       & \cmark   & \cmark   & \cmark & \xmark & \xmark \\
Instruction retriever~{\scriptsize\citep{weller2024promptriever}}       & $\sim$   & $\sim$   & \cmark & \xmark & $\sim$ \\
Fine-tuned ColBERT                                                      & \cmark   & \cmark   & \xmark & \cmark & \xmark \\
\oursrow \textbf{\excise{}} (ours)                                  & \cmark   & \cmark   & \cmark & \cmark & $\sim$ \\
\bottomrule
\end{tabular}}
\end{table}

\paragraph{Neural retrieval architectures.} Modern retrievers fall into three families. Single-vector dense
retrievers encode a query and a document into one embedding each and rank by inner product, as in
DPR~\citep{karpukhin2020dpr}, ANCE~\citep{xiong2021ance} and Contriever~\citep{izacard2022contriever}, and in
the widely used E5~\citep{wang2022e5}, BGE~\citep{xiao2023bge} and GTE~\citep{li2023gte} embedding models.
Sparse lexical models such as SPLADE~\citep{formal2021splade} retain interpretable term weights in the
tradition of BM25~\citep{robertson2009bm25}. Late-interaction models keep a separate embedding per token and
score by MaxSim, an approach introduced by ColBERT~\citep{khattab2020colbert} and refined in
ColBERTv2~\citep{santhanam2022colbertv2} and PLAID~\citep{santhanam2022plaid}.

\excise{} builds on late interaction. The underlying problem is general, since any score that rewards
query--document similarity also rewards the excluded tokens. The identity of Proposition~\ref{prop:mono},
however, is specific to MaxSim, which is why we state the diagnosis and the repair for late interaction rather
than for retrieval in general.

\paragraph{Negation benchmarks.} NevIR~\citep{weller2024nevir} pairs two documents that differ only in a
negation and asks a retriever to prefer the one that the query describes. The negation is in the document, the
candidate set has two members, and the metric is pairwise accuracy. Exclusion as we study it differs in each
respect: the negation is in the \emph{query}, the candidate set is the corpus, and the system must keep a
document out of the top ten instead of ordering two documents correctly. ExcluIR~\citep{zhang2024excluir}
moves the negation to the query and reports large drops for dense models; this is the setting we adopt. The
SIGIR reproduction of NevIR~\citep{vandenelsen2025reproducing} fine-tunes several architectures on one
benchmark and evaluates them on the other. It finds that only the cross-encoder transfers its negation
ability, whereas listwise LLM rerankers learn the source benchmark comparably well but do not generalise to
the target. This result may appear to conflict with Table~\ref{tab:strong}, where off-the-shelf cross-encoders
solve few exclusion queries, but the two are compatible. The reproduction uses pairwise accuracy, and a model
can order two documents correctly without suppressing either. Success@10 additionally requires that no
excluded document remains in the top ten over a full corpus. A cross-encoder that has learned to prefer the
correct document of a pair has not thereby learned to remove the incorrect one from a ranking, and
Table~\ref{tab:strong} separates these two abilities.

\paragraph{Lexical and Boolean filtering.} The classical approach to an exclusion query is lexical: detect
the negation cue, then delete or down-weight the terms it governs. This approach has two limitations. It
requires an explicit cue, which the $569$ cue-free queries do not contain (Appendix~\ref{app:human}). It also
operates on surface forms, so a document that names the excluded topic by a synonym is not affected. The
closest method measured in this paper is the sparse BM25 demotion of Appendix~\ref{app:ninefix}, which is a
stronger baseline than a cue-triggered filter because it \emph{receives} the excluded topic instead of
extracting it. Even so, it requires a sparse index maintained alongside the dense one, and it falls from
$0.8520$ to $0.8159$ when the topic comes from a detector instead of the record.

\paragraph{Instruction-following retrieval.} Exclusion is one instance of following an instruction attached to
a query. TART~\citep{asai2023tart} and Instructor~\citep{su2023instructor} prepend task instructions during
training; FollowIR~\citep{weller2024followir} and InstructIR~\citep{oh2024instructir} benchmark instruction
adherence; and Promptriever~\citep{weller2024promptriever} trains a retriever that can be prompted like a
language model. These systems are capable but costly: they build on large instruction-tuned encoders, require
the instruction at indexing or query time, and give no \emph{structural} guarantee that following one
instruction leaves unrelated queries unaffected. On ordinary retrieval, they differ from \excise{} in this guarantee and not necessarily in the measured outcome. Fine-tuned Promptriever is the one external system we measure that maintains its
ordinary-retrieval level, at $+0.033$ nDCG@10, so the trade-off reported in this paper is not universal. It
does not reach our exclusion levels, however, and trails \excise{} on all six collections
(\S\ref{sec:results-strong}). \excise{} isolates the single capability of exclusion in a small query-time
module and returns the frozen ranking unchanged whenever its detector does not fire.

\paragraph{Rerankers and query-time intervention.} A cross-encoder reranker~\citep{nogueira2019passage,
nogueira2020monot5, zhuang2023rankt5} reads a query and a document jointly and could in principle attend to a
negation. Two structural costs nevertheless remain. It re-scores every candidate with a full joint forward
pass, whether or not the query contains an exclusion. It is also trained for relevance rather than
suppression, so it has no component dedicated to removing a named topic and offers no guarantee for ordinary
queries. Our measurements are consistent with both points: fine-tuning the four rerankers on exclusion costs
them $0.043$--$0.095$ nDCG@10 on suites that contain no exclusions. A different line of work places the logic
outside the encoder: neural-symbolic retrieval~\citep{xu2025nsir} maps queries and documents into first-order
logic and re-ranks by logical consistency. This addresses negative constraints directly, but it adds a symbolic
stage to the query path and does not leave the ranking unchanged when no constraint is present. \excise{}
shares the retrieve-then-rescore structure but replaces the joint model with a small query-side detector, an exclusion adapter and a parameter-free demotion rule. It leaves the ranking unchanged when its detector does not fire, and it never modifies the index.

\paragraph{Relevance feedback.} Negative relevance feedback is the oldest method for moving a ranking away
from unwanted content: Rocchio~\citep{rocchio1971relevance} moves the query vector away from the centroid of
documents marked non-relevant. Both Rocchio and \excise{} must be told what to avoid, and they differ in the
form of that information. Rocchio requires a set of \emph{documents}, marked by a user or collected in a
previous round, which an exclusion query does not provide. \excise{} requires a \emph{span}, which the query
already contains, so a single pass is sufficient. We evaluate the centroid form as one of the nine test-time
interventions in Appendix~\ref{app:ninefix}, where, like the others, it must be given the topic.

\paragraph{Parameter-efficient adaptation.} Adapters and LoRA~\citep{houlsby2019adapters,hu2022lora}
train a small number of new parameters while the rest of the model is frozen. Applied on the document side,
they are inexpensive to train but not to deploy, because the corpus must be re-encoded and the index rebuilt at
every model update, as for a full fine-tune. The components of \excise{} are query-side LoRAs, so the
contribution is not the adapter itself but the finding that a component of this kind, gated by a switch and confined to
the query path, adds exclusion without re-encoding the corpus.

\paragraph{Domain and conversational search.} Prior work studies question answering and information extraction over sustainability and regulatory documents~\citep{ali2026sustainableqa,ali2026regreact}, as well as multi-turn conversational search~\citep{ali2026recor,ali2026memrevise}. \xbench{} covers the first setting through its ESGenius, EDGAR and EUR-Lex collections, but contains only single-turn queries.

\end{document}